\documentclass[journal,comsoc]{IEEEtran}
\usepackage[table]{xcolor}
\usepackage{algorithm}
\usepackage{algpseudocode}
\usepackage{amsmath}

\usepackage{graphicx} 
\usepackage{multicol}
\usepackage{epstopdf}
\usepackage{cite}
\usepackage{float}

\usepackage{enumitem}
\usepackage{array}
\usepackage{xcolor}
\usepackage{lipsum}
\usepackage{fixltx2e}
\usepackage{amsfonts}
\usepackage{algpseudocode}%
\usepackage[numbers,square, comma, sort & compress]{natbib}
\ifCLASSOPTIONcompsoc
\usepackage[caption=false,font=normalsize,labelfont=sf,textfont=sf]{subfig}
\else
\usepackage[caption=false,font=footnotesize]{subfig}
\fi

\ifCLASSINFOpdf
\else
\fi
\usepackage{amsmath}
\usepackage[cmintegrals]{newtxmath}
\usepackage[T1]{fontenc}
\usepackage{lipsum}
\usepackage{bbm}
\usepackage{caption}
\usepackage{algorithm}
\usepackage{algpseudocode}
\usepackage{mathtools}

\newtheorem{thm}{Theorem}
\newtheorem{lemma}{Lemma}

\usepackage{dblfloatfix}
\usepackage{array}
\ifCLASSINFOpdf
\else
\fi
\usepackage{amsmath}
\usepackage[cmintegrals]{newtxmath}
\ifCLASSOPTIONcompsoc
\usepackage[caption=false,font=footnotesize,labelfont=sf,textfont=sf]{subfig}
\else
\usepackage[caption=false,font=footnotesize]{subfig}
\fi
\usepackage{bbm}

\usepackage{algpseudocode}

\begin{document}

\title{Multi-player Multi-armed Bandits for Stable Allocation in Heterogeneous Ad-Hoc Networks}
\author{Sumit J. Darak and Manjesh K. Hanawal
	\thanks{Sumit J. Darak is with the Electronics and Communications Department, IIIT Delhi, India.}
	\thanks{Manjesh K. Hanawal is with Industrial Engineering and Operations Research Department, IIT Bombay, India.}}

\maketitle

\begin{abstract}
	Next generation networks are expected to be ultra-dense and aim to explore spectrum sharing paradigm that allows users to communicate in licensed, shared as well as unlicensed spectrum. Such ultra-dense networks will incur significant signaling load at base stations leading to a negative effect on spectrum and energy efficiency. To minimize signaling overhead, an ad-hoc approach is being considered for users communicating in the unlicensed and shared spectrums. For such users, decisions need to be completely decentralized as: 1) No communication between users and signaling from the base station is possible which necessitates independent channel selection at each user. A collision occurs when multiple users transmit simultaneously on the same channel, 2)~Channel qualities may be heterogeneous, i.e.,  they are not same across all users, and moreover, are unknown, and 3)~The network could be dynamic where users can enter or leave anytime. We develop a multi-armed bandit based distributed algorithm for static networks and extend it for the dynamic networks. The algorithms aim to achieve stable orthogonal allocation (SOC) in finite time and meet the above three constraints with two novel characteristics: 1)~\textcolor{black}{Low complexity} narrowband radio compared to wideband radio in existing works, and 2)~Epoch-less approach for dynamic networks. We establish convergence of our algorithms to SOC and validate via extensive simulation experiments.
	 
\end{abstract}


\section{Introduction}
Next generation wireless networks such as 5G aim to offer the wide range of new services such as enhanced local broadband, high-speed multimedia, mission-critical control, private networks such as Industrial IoT and enterprise \cite{NR} via spectrum sharing. Such networks with diverse service requirements are expected to greatly enhance user experience \cite{NR}.
Recently, 3GPP proposed a new radio (NR) based heterogeneous networks consisting of base stations of various sizes. Compared to existing networks, NRs can operate not only in licensed spectrum but also in the shared (2.3GHz/ 3.5GHz) as well as unlicensed spectrums (2.4GHz/ 5-7GHz/ 57-71GHz). Such network opens up many interesting challenges such as resource allocation, dynamic and context-aware network adaptation, and in-depth knowledge discovery in the complex environment for which machine learning and artificial intelligence frameworks offer novel solutions \cite{NR,ca1,ca2,p2}.
 
The next generation networks are envisioned to work on the principle of separate signaling (large base station) and data infrastructure  (small base stations) which allows adaptation of data network to the current traffic situation while maintaining the coverage.  These networks will be ultra-dense with very high peak rate but relatively lower expected traffic per network node \cite{NR}. This makes signaling (control communications) component to be a substantial part of the network traffic leading to a negative effect on the energy and spectrum efficiency.  Further, separate signaling and data infrastructures put a significant signaling load on the large base stations, especially in ultra-dense networks. To reduce this, the ad-hoc network approach is being considered for users utilizing unlicensed and shared spectrum \cite{NR,ca1,ca2,p2}. This not only reduces the signaling load at base stations but also allows the higher number of users per base station (dense networks). However, the channel selection needs to be done independently at each user since the ad-hoc network does not support any direct communication/coordination between users. In addition, channel statistics may not be known which requires it to be learned and further channel statistics may differ across the users. In this paper, we explore multi-player multi-armed bandit (MPMAB) framework which enables learning and coordination tasks at each user thereby improving the spectrum and energy efficiency of the ad-hoc networks. 

 \textcolor{black}{MPMAB is a variant of the stochastic multi-armed bandits (MAB) where all players/users aim to maximize the total throughput in a distributed fashion by selecting a set of common arms/channels.}
Due to hardware and power constraints of battery operated users, we assume that a user can either sense or transmit but cannot do both simultaneously. Furthermore, a user can sense or transmit only over a single channel in each time slot. Such radio terminals are referred to as narrowband radios. In ad-hoc networks, the users cannot communicate/coordinate with each other and may not know the number of other users in the network. If two or more users transmit on the same channel simultaneously, they experience `collision'  and the packet needs to be transmitted again. Such collisions results in throughput loss and also do not provide any information about the channel status. In addition to unknown statistics, channels are heterogeneous where the average throughput on a channel may not be the same for all users. Later, we consider the dynamic networks where users may enter or leave without prior agreements. Even though all users employ the same algorithm, new users need to learn to coordinate without any prior knowledge of the status of the other users in the network. Such task poses a real challenge in the ad-hoc networks and is the focus of this paper. 

We develop distributed algorithms for static and dynamic ad-hoc networks that enable users to reach a stable orthogonal configuration (SOC). Under SOC no two users would simultaneously improve their throughput if they swap their channels. Reaching an SOC quickly is critical as it allows the users to transmit on one of their preferred channels without incurring a significant number of collisions. Our contributions can be summarized as follows: 

\begin{itemize}
	\item  For a static ad-hoc network with an unknown number of users and heterogeneous channels, we develop the dSOC\_SN algorithm which converges in finite time to a SOC with high probability. \textcolor{black}{The convergence rate is of order $O(1/\Delta_{\min}^2 + K^3 \log (1/\delta))$, where $\Delta_{\min}$ is a problem dependent constant and $\delta$ is the confidence parameter.}
	\item  For a dynamic ad-hoc network with heterogeneous channels and unknown number of users, we develop the dSOC\_DN algorithm. We give a high probability upper bound on the time to reach a SOC after a user leaves or enters into the network. The dSOC\_DN algorithm is the first algorithm for the dynamic heterogeneous networks and is based on the novel epoch-less approach (without restarting the algorithm).  
	\item We validate the performance of both dSOC\_SN and dSOC\_DN algorithms through extensive simulations.  They outperform the state-of-the-art algorithm in \cite{lax} for heterogeneous ad-hoc networks. 
	\item Our algorithms need low complexity single antenna narrowband radio compared to wideband radios in existing works such as \cite{lax}. Unlike \cite{MC,wiopt,wcnc}, the dSOC\_DN algorithm does not require global clock synchronization for new users thereby further reducing the algorithm and radio complexity.
	 These advantages make our algorithms suitable for decentralized and battery operated networks.
\end{itemize}
 
The paper is organized as follows. The related work and network model are discussed in Section~\ref{sec:RW} and Section~\ref{sec:NM}, respectively. The proposed algorithms and their analysis are presented in Section~\ref{sec:SN} and Section~\ref{sec:DN} for static and dynamic networks, respectively. The simulation results are presented in Section~\ref{sec:exp} and Section~\ref{conclusion} concludes the paper.


\section{Related Work}\label{sec:RW}
\begin{table*}[!b]
	\centering
	\caption{Comparison of Various Distributed Algorithms}
	\label{table1}
	\resizebox{\textwidth}{!}{%
		\color{black}\begin{tabular}{|l|l|l|l|l|l|}
			\hline
			\textbf{Algorithms} & \textbf{Channel Model} & \textbf{Network Model} & \textbf{Radio Model} & \textbf{Channel} & \textbf{Additional} \\
			&  &  &  & \textbf{Allocation} & \textbf{Assumptions} \\\hline
			\textbf{$\rho^{rand}$, MCTopM \cite{zhandi,gai2,prand,emilie,lugosi}} & Homogeneous & Static & Narrowband & Optimal & \begin{tabular}[c]{@{}l@{}}$N$ known,\\ $N\leq K$\end{tabular} \\ \hline
			\textbf{MC, MEGA, SCF, TSN/TDN \cite{MEGA,MC,wiopt,wcnc}} & Homogeneous & Static and Dynamic & Narrowband & Optimal & \begin{tabular}[c]{@{}l@{}}Bound on sub-optimality gap ($\Delta$) known,\\ New user knows network status\\ $N\leq K$\\ Epoch except TDN\end{tabular} \\ \hline
			\textbf{dE3 \cite{rjain}} &  Heterogeneous & Static & Narrowband & Optimal & \begin{tabular}[c]{@{}l@{}}players execute Bersekas auction,\\ $N\leq K$\end{tabular} \\ \hline
			\textbf{M-ETC and ESER \cite{OptimalAllocation, M-ETC}} &  Heterogeneous & Static and Dynamic & Narrowband & Optimal &  \begin{tabular}[c]{@{}l@{}}New user knows network status\\ $N\leq K$\end{tabular} \\ \hline
			\textbf{CSM\_MAB \cite{lax}} & Heterogeneous & Static & Wideband & Stable & $N\leq K$ \\ \hline
			\textbf{This work} & Heterogeneous & Static and Dynamic & Narrowband & Stable &  \\ \hline
		\end{tabular}%
	}
\end{table*}
Various works dealing with the coordination in the multi-user ad-hoc networks have been discussed in the literature. In this section, we focus on works employing the multi-armed bandit (MAB) based approach for channel selection 
\textcolor{black}{The MAB approach provides a standard framework for learning in uncertain environments \cite{MAB1,MAB2} and is applied extensively in the study of centralized as well as ad hoc networks.} \textcolor{black}{In this paper, we focus on the multi-player extension of MAB framework where multiple players play the same set of arms. Here the algorithms are augmented with the distributed signaling schemes that aim to achieve optimal sum reward for the players without requiring a central coordinator or a control channel.} We use the $N$ and $K$ to denote the number of users and channels, respectively.

$\rho^{rand}$ \cite{prand} is one of the first distributed algorithms for a static homogeneous ad-hoc network with known number of users. It combines the well known upper confidence bound (UCB) with rank based randomization approach to orthogonalize users in the best $N$ channels. Though $\rho^{rand}$ offers asymptotic logarithmic regret (throughput loss), it incurs a large number of collisions in the process. Subsequent algorithms in \cite{zhandi,gai2,emilie,lugosi} are based on $\rho^{rand}$, and offer further improvement in performance by reducing the number of collisions.
MCTopM algorithm in \cite{emilie} achieves faster orthogonalization with a better back-off mechanism on collisions. Recent works in \cite{lugosi, SIC_MMAB} consider a restricted feedback setup where the users cannot observe collisions and manage to develop algorithms with logarithm regret. The works in \cite{wiopt,wcnc} consider distributed channel assignment in ad-hoc networks where users opportunistically transmit on licensed channels. \textcolor{black}{The major drawbacks of these algorithms \cite{zhandi,gai2,prand,emilie,lugosi,SIC_MMAB, wiopt,wcnc} are that they need prior knowledge of $N$, require network to be static and homogeneous. Such assumptions are unrealistic in ad-hoc networks and they do not extend to the cases when any of these assumptions do not hold.}

Recently, attempts are made to study both static and dynamic netwroks in homogeneous setting \cite{MEGA,MC,TDN,quek} have been proposed to overcome these drawbacks. The MEGA algorithm \cite{MEGA} uses the classical $\epsilon$-greedy MAB algorithm and ALOHA based collision avoidance mechanism. Though collision frequency reduces in MEGA as the game proceeds, it may not go to zero as shown in \cite{MC}. To overcome this \cite{MC} develops MC algorithm that incurs collisions due to random hopping (RH) in the initial learning phase and guarantees collision-free access over optimum channels subsequently. Though MC performs better than MEGA, its performance in the learning phase is poor -- MC forces a large number of collisions to get a good estimate of $N$. TSN and TDN algorithms in \cite{TDN} overcome these issues by using `Trekking' approach where users orthogonalize on top $N$ arms without estimating $N$. 


\textcolor{black}{The algorithms in \cite{rjain, OptimalAllocation, M-ETC, lax, leshem} work for the heterogeneous channels. They force collisions to exchange information between the users. Specifically, the players collide with others in a specific pattern to convey their estimate of rewards or preference of channels.   
The dE3 algorithm in \cite{rjain} employs Bertsekas auction algorithm which requires the users to exchange bids to win the channel of their preference via collisions. ESER \cite{OptimalAllocation} and M-ETC \cite{M-ETC} algorithms allow users to exchanges the mean values they observe with others via collisions. These algorithms achieve near-logarithmic regret (logarithm when the optimal allocation is unique). However the main drawback of dE3, M-ETC, ESER is that they require frequent exchange of information between the users resulting in significant throughput loss due to signal overheads}. Coordinated Stable Marriage MAB (CSM\_MAB) algorithm in \cite{lax} overcomes the need for frequent communications by aiming to achieve stable allocation rather than optimal allocation. However, it requires that all users to simultaneously sense all channels (wideband sensing) which is,  as discussed later in this section, computationally complex and expensive and hence may not be suitable for low cost, battery operated user terminals. Our goal in this work is to achieve stable allocation as in \cite{lax} in a heterogeneous network with minimum signaling overhead and using simpler narrowband radios that are well suited for ad-hoc networks with power constraints.  

\textcolor{black}{Our contributions in this work is to design distributed algorithms for heterogeneous channels for both static and dynamic network scenario. Furthermore, in our work each user can sense and transmit over only one channel in each time slot (narrowband sensing) which is more realistic and computationally efficient than wideband sensing in \cite{lax}. Our novel block structure makes the protocol simple, easy to implement and achieves performance better than that achieved by CSM\_MAB using wideband sensing \cite{lax}. It also ensures the stability of the network when $N\geq K$ as discussed in Section~\ref{NgeK} and also overcomes the need of global clock synchronization which is a significant advantage for the dynamic networks.}\vspace{-0.2cm}

\subsection{Radio Models}
One of the major aspects of the distributed algorithm is the capability of a radio terminal. Existing distributed algorithms consider various types of radio architectures which not only impact the learning period but also the complexity and performance of the distributed algorithm. These architectures offer a trade-off between sensing/transmission capability and implementation complexity. For instance, \cite{lax} considers sophisticated radio terminals with two independent analog signal processing (ASP) blocks each consisting of an antenna, matching units, amplifiers, analog-to-digital or digital-to-analog converters, etc. One ASP block is used for narrowband (single channel) transmission while second ASP block can sense all channels simultaneously, i.e. wideband sensing.  Such wideband sensing makes an estimation of $N$ trivial and simplifies coordination since users can differentiate between users on different channels. However, the wideband channel sensing needs high-speed ADCs making ASP as well as subsequent digital baseband processing complex and power hungry, and hence not suitable for battery operated radio terminals \cite{nus}. The non-contiguous wideband channel sensing is even more challenging. Another architecture consisting of two narrow-band ASP blocks which allow simultaneous transmission and sensing over different channels have been considered in \cite{rjain}. In this paper, we consider the architecture which has the lowest complexity among the three. It consists of a single narrowband ASP chain which allows either transmission or sensing over a single channel in a given time slot. Such architecture can detect the presence of another user on their channel either by experiencing collision or sensing but cannot estimate the number of collided or sensed users. Furthermore, the architecture cannot sense multiple channels simultaneously making the orthogonalization and establishing coordination extremely challenging than in \cite{rjain,lax}. In Table I, we compare existing distributed algorithms with respect to various parameters.


\section{Network Model}\label{sec:NM}

Consider an ad-hoc network consisting of $N$ users competing for $K (\geq N)$ channels in an unlicensed spectrum. We assume the communication is time slotted, and the users are clock synchronized with respect to the beginning of each time slot as in \cite{MEGA,MC,wiopt,wcnc,rjain,lax,prand,zhandi,gai2,quek}. In each time slot, each user can transmit only once over any one of the $K$ channels. When two or more users transmit simultaneously on a channel, a collision occurs and all the \textcolor{black}{users} involved in the collision need to re-transmit the lost packet. The users are not aware of how many other users are present ($N$ is unknown) and no central coordinator exists to facilitate their channel selections.  



Another major characteristic of our network is that the channels are heterogeneous, i.e., the expected reward/throughput on a channel depends not only on the channel but also on the user selecting it. Such model is more practical than homogeneous channels in \cite{MC,wiopt,wcnc} as it considers location-dependent channel conditions. Let $\mu_{n,k} \in [0 \; 1]$ denote the expected reward for user $n \in [N]$ on channel $ k \in [K]$ on a collision-free transmission. These mean values are unknown to the users and user $n$ can only observe $\{\mu_{n,k}, k\in [N]\}$, i.e., all observations are local and a user does not know the expected reward offered by the channels to other users. The reward observed by a user on a channel under collision-free transmissions is assumed to be independently and identically distributed. The same setup is also considered in \cite{rjain,lax}. 

The performance of a distributed algorithm is compared in terms of expected rewards/throughput. The maximum reward is achieved when all users are on orthogonal channels and channel allocation guarantees maximization of the sum of rewards over all users. Formally, let $\pi: [N]\rightarrow [K]$ denotes an orthogonal allocation of the users to the channels and $\mathcal{C}$ denotes all such possible allocations. Then, the maximum expected reward is given by
\begin{equation}
R_{max} = \max_{\pi \in \mathcal{C}} \sum_{n=1}^{N} \mu_{n,\pi(n)}.
\end{equation}
To achieve $R_{max}$ each user need to know the expected reward of all channels for all users. This requires all the users to share their observation with all other users in the networks which either need direct communication between users or sophisticated signaling scheme. Instead, we focus on achieving a stable orthogonal allocation configuration (SOC) \cite{lax} where no two users can simultaneously agree to swap their channels without one of them getting a `less preferred' channel than its current one.  

To explain SOC, we first define the rank of a user which corresponds to the number of channels whose expected reward is higher than the current channel the user has selected.  For the $n$-th player, \textcolor{black}{its value in the $t$ round} is given by,
\begin{equation}
\gamma_n(t) = \sum_{k=1}^{K} \textbf{1}\{\mu_{n,k}> \mu_{n,\pi_t(n)}\},
\end{equation}
where $\pi_t(n)$ indicates the channel selected by user $n$ in time slot $t$ using policy $\pi:=\{\pi_t: t\geq1\}$. The total rank of the network (also called network potential) is given by
\begin{equation}
\gamma_{\pi}(t) = \sum_{n=1}^{N} \gamma_n(t).
\end{equation}

An assignment $\pi$ is said to be SOC if a swap of channels between any pair of users or switch to the vacant channel do not strictly decrease network potential. For a network in SOC, the user will have no incentive to request any other user for a swap of their channels, hence channel switches/swaps will not occur. Our aim is to design distributed algorithms that converge to a SOC as quickly possible. We note that there could exist multiple SOCs.

\section{Static Network: Algorithm and Analysis}\label{sec:SN}
In this section, we consider the static network where all users simultaneously enter into the network at the beginning ($t=0$) and remain active until the end. We describe an algorithm named distributed Stable Orthogonal Configuration for Static Network ($dSOC\_SN$) and analyze its performance.

\subsection{dSOC\_SN Algorithm}
The algorithm comprises of two phases: 1)~Random hopping (RH), and 2)~Sequential master and channel switching (SMCS). The pseudo code is given in Algorithm~\ref{Algo1}, where $K$ indicates the number of channels. The same algorithm is run independently at each user. For the $n$th user $\pi_{T_{rh}}(n)$ indicates the channel on which user $n$ gets locked in the RH phase.

\begin{algorithm}[!h]
	\caption{dSOC\_SN Algorithm} \label{Algo1}
	\begin{algorithmic}
		\State	Input: {$K$}
		\State	$\pi_{T_{rh}}(n)= RH(K)$ 
		\State	 SMCS $(\pi_{T_{rh}}(n),K,1)$ 
	\end{algorithmic}
\end{algorithm}

\subsubsection{Random Hopping (RH) Phase}
The RH phase allows users to orthogonalize on different channels. This phase is an adaption of the RH phase in \cite{wiopt} where licensed spectrum is considered as opposed to unlicensed spectrum here. The pseudo code of the RH phase is given in Subroutine $1$. In RH phase, each user selects a channel drawn uniformly at random (line $7$) in each time slot. Once the user observes a collision-free transmission on a channel, user {\it locks} on that channel and we refer to it as the reserved channel of the player.    The duration of the RH phase is fixed and is set equal to  $T_{rh}$ (See Lemma \ref{lma:RanHop_Static}). Each player can can lock on its reserved channel at any slot within RH phase, and once locked, she plays her reserved channels in the rest  RH phase. If user does not get reserved channel within $T_{rh}$ slots, she has to leave the network. 

\begin{algorithm}[!h]
	\caption*{\textbf{Subroutine 1:} Random Hopping (RH)}
	\begin{algorithmic}[1]
		\State Input: $K$  
		\State Set $Lock=0$ and compute $T_{rh}$ using Eq.~\ref{eq:trh}.
		\For{$t=1 \dots T_{rh}$}
		\If{($Lock == 1$)}
		\State Choose channel, $\pi_n(t) = \pi_n(t-1)$
		\Else
		\State Randomly choose channel, $\pi_n(t) \sim U(1, ... ,K)$ 
		\State Set $Lock=1$ if no collision is observed.
		\EndIf
		\EndFor	
		\State Return $\pi_{n}(T_{rh})$ indicating the channel index at $t=T_{rh}$.
	\end{algorithmic}
\end{algorithm}

\subsubsection{Sequential Master and Channel Switching (SMCS) Phase}
All users enter into the SMCS phase at time $t=T_{rh}+1$ and continue in that phase till the end. In this phase, the users maintain a list of their preferred channels, i.e., the channels that are better than their current one, and taking turns, request other users who are currently on one of their preferred channels to switch their channel with them. If a preferred channel happens to be unoccupied, it is simply taken, otherwise, the user gets it only if the other user accepts the switch request. The other user accepts the request only if the channel she will be shifting to is also one of her preferred channels, otherwise, she rejects the request. If the switch request is rejected, the user tries her next preferred channel. To allow such negotiations, we divide the time slots into blocks which further consists of sub-blocks. \textcolor{black}{The SMCS phase involves two tasks: 1) Statistic learning to rank the channels as per their throughput, and 2) Channel switching. We first discuss the novel signaling scheme which allows users to switch their channel without direct communication followed by MAB algorithm based statistic learning.}

\subsection{Signaling for Channel Switching}

\begin{figure}[!h]
	\centering
	\vspace{-0.1cm}
	\includegraphics[scale=0.4]{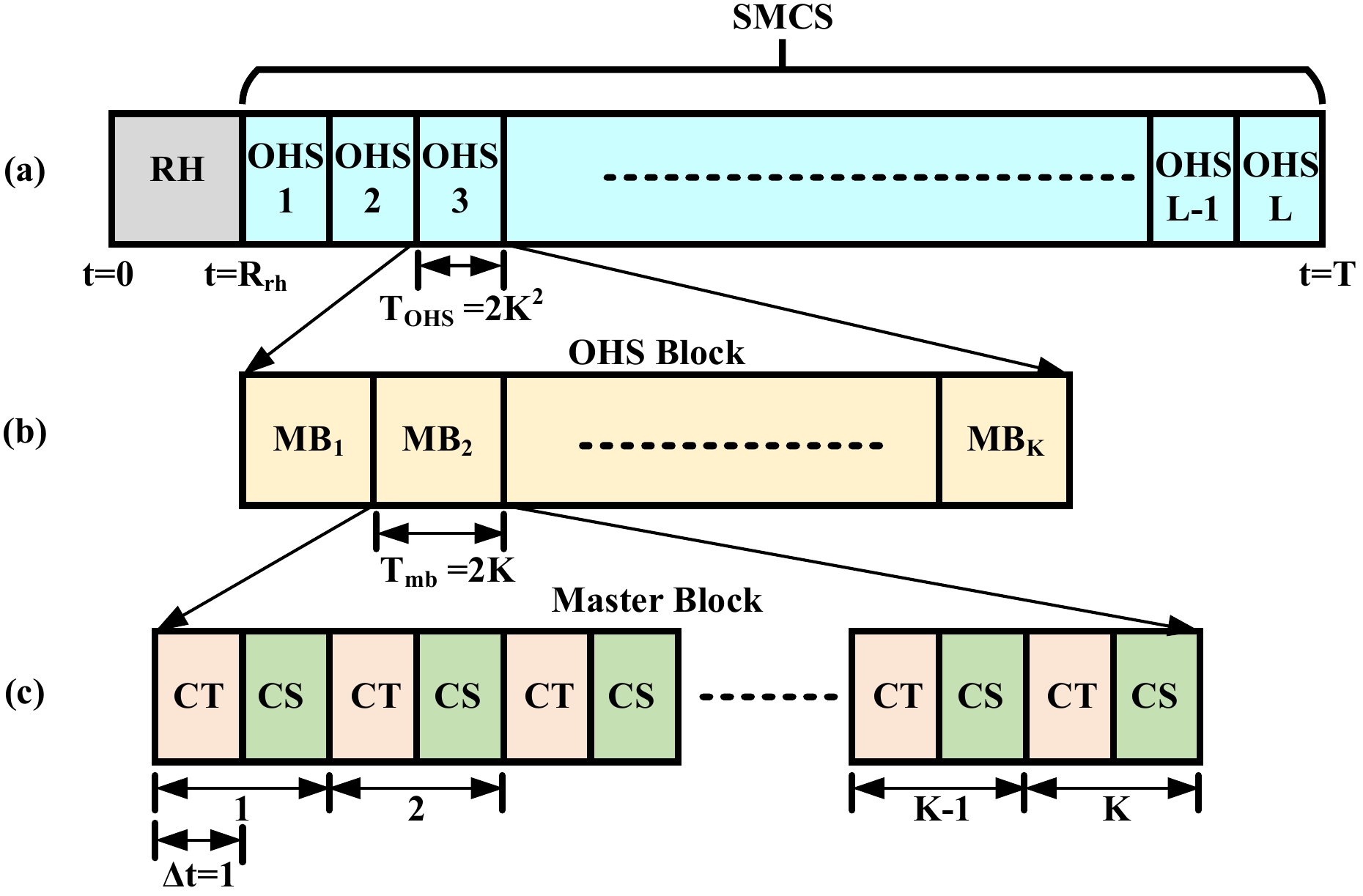}%
	\caption{\footnotesize{Different phases and sub-blocks of the dSOC\_SN algorithm for static ad-hoc network.}}
	\label{fig:static_alg_td}
		
\end{figure}

The SMCS phase consists of a sequence of one hot switching (OHS) blocks each of $T_{ohs}$ time slots that repeat one after another. Each OHS block consists of $K$ master blocks (MB) and each master block is made up of $T_{mb}$ time slots. Fig.~\ref{fig:static_alg_td} gives the structure of the OHS and master blocks. The duration of each OHS block is $T_{ohs}= K T_{mb}$ slots.  

Each user uses the channels they get at the end of the RH phase as their {\it reserved} channel and \textcolor{black}{attempts to get a channel that is currently being ranked higher than her current reserved channel by its own MAB algorithm (discussed in next sub-section).} This is achieved by allowing each user to become a {\it master} in a specific MB. At any time, only one user is allowed to be a master while all other users continue to transmit on their respective reserved channels and are referred to as {\it non-masters}.  When a user becomes a
\textit{master}, she requests other users for a switch of channels that are better than her current channel. If she gets a better channel,
she continues to transmit on it till she becomes the master again. If she moves to a new channel, it becomes her reserved channel, otherwise, the current channel will be her reserved channel.

\begin{figure}[!h]
	\centering
	\includegraphics[scale=0.32]{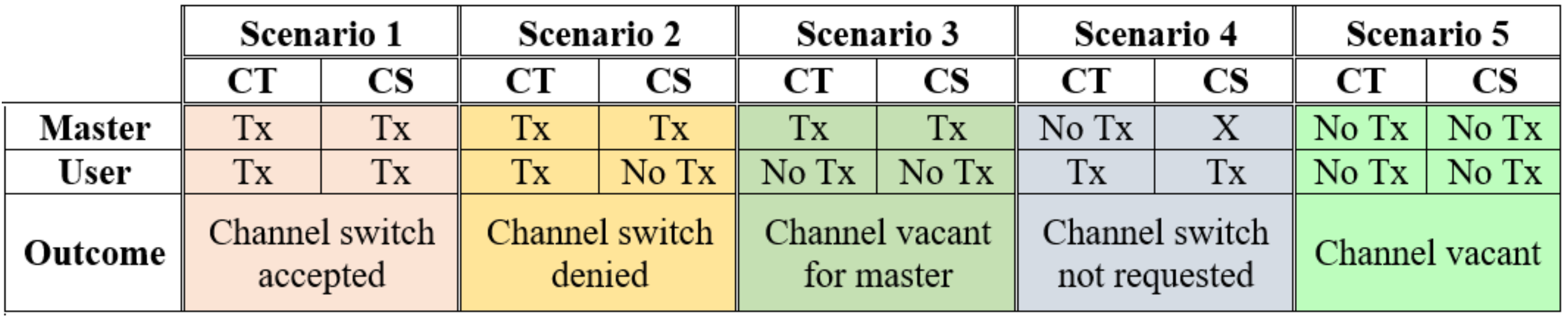}%
	\vspace{-0.1cm}
	\caption{\footnotesize{Various channel switching scenarios between master and non-masters in a given CT/CS time slots. Note that Tx indicates transmission and X indicates don't care.}}
	\vspace{-0.5cm}
	\label{fig:ST_SC}
\end{figure}

In the $k^{th}$ MB, the user on the $k^{th}$ channel becomes the {\it master} and gets a chance to move to a better channel by sending switching requests on channels that are better than her current channels. 
This is felicitated by a signaling scheme defined as follows:
The MB block consists of $K$ sub-blocks (SB) of two slots each (See Fig.~\ref{fig:static_alg_td}(c)). The first slot in each SB is referred to as channel transmit (CT) and the second as channel switch (CS). In the $i-$th SB, master selects her $(i-1)-$th preferred channel\footnote{$0$-th preferred channel corresponds to her current reserved channel} and transmits. If a no collision is incurred in the CT slot, then it implies that no user is present on that channel and the master switches to it and makes it her reserved channel (scenario $3$). If a collision is observed in a CT slot, it implies that another user is present on that channel and the master transmits on the same channel in the next CS slot.  If the master does not encounter a collision in the CS slot after the collision in previous CT slot, it implies the current user on that channel is not willing to exchange her channel and rejected master's request to switch channels (scenario $2$). If the master encounters a collision in the CS slot, it implies that the user on the current channel accepts master's request for channel switch (scenario $1$), in which case the master switches to the requested channel and makes it her new reserved channel. These and other possible scenarios are given in Fig~\ref{fig:ST_SC}.  If the master gets a new channel no more switches are requested in the current MB, otherwise the process is repeated in the next SB for the next preferred channel.

\subsection{Learning the Channel Statistics} 
\textcolor{black}{As discussed in Section~\ref{sec:NM}, users are not aware of the expected reward/throughput they receive on each channel and need to learn and index (or rank) them. We employ the multi-armed bandit (MAB) approach based learning algorithm to find the index for each channel. Various indexing methods based on Upper Confidence Bound (UCB), Bayes-UCB algorithm, Thompson Sampling (TS)  can be used \cite{kaufmann}. In this paper, we focus on indexing based on UCB and its analysis. However,  our algorithm can work with other indexing methods as well. The UCB algorithm is based on exploration-exploitation trade-off and its indexing is based on the optimistic estimates of mean rewards. For the $n$ user, the channels are indexed  based on the UCB scores given as \cite{MAB1,MAB2},
\begin{equation}
\label{eqn:UCBInd}
Q_{n,k}(t)= \frac{P_{n,k}(t)}{S_{n,k}(t)} + \sqrt{\frac{2 \log t}{S_{n,k}(t)}} \quad \forall k \in [K],
\end{equation}
where $P_{n,k}$ is the total reward received by user $n$ on channel $k$ when it was chosen for $S_{n,k}$ time slots. The UCB algorithm is asymptotically optimal in the sense that it selects the sub-optimal channels an exponentially smaller number of times compared to the optimal channel.}

\subsection{Duration of Master Block}

Note that the block structure allows the users to identify the reserved channel of the master in each MB which helps them to decide whether to accept or reject a request to switch. Specifically, if a user receives a switch request in the $k$-th MB, then she knows that she will move to $k$-th channel on accepting the request.

Since there are $K$ channels, a master can switch to any one of the ($K-1$) channels which means the duration of MB can be set at most $2(K-1)$ slots. However, we force the master to transmit on its reserved channel in the first SB and thereafter follows its preference list in the rest of the ($K-1$) SB. Hence, the duration of MB is $T_{mb} = 2K$ slots. As we will see later in Section~\ref{sec:DN}, this block structure allows new users to synchronize with the existing users in the dynamic networks easily.

\begin{algorithm}[!b]
	\caption*{\textbf{Subroutine 2:} SMCS Phase}
	\begin{algorithmic}[1]
		\State Input: $\pi_{n,r},K,  M_{ind}$
		\For {$OHS$=1,2,....}
		\For {$MB= M_{ind},2,..,K$} \Comment{master block counter}
		\If {$\pi_{n,r}==MB$} 
		\State Enter into \textbf{Master} mode.
		\Else
		\State Enter into \textbf{non-Master} mode.
		\EndIf		
		\EndFor
		
		\EndFor
	\end{algorithmic}
\end{algorithm}

\begin{algorithm}[!b]
	\caption*{\textbf{Subroutine 3:} Master Mode}
	\begin{algorithmic}[1]
		\State Input: $\pi_{n,r},K$
		\State Transmit on $\pi_{n,r}$ for two time slots. \Comment{First SB}
		\For {$SB$=1,2,....,K-1}
		\State Transmit over $SB^{th}$ channel in the preference list for \quad two slots (CT and CS).
		\If {No Collision on  $\textit{CT}$ slot} \Comment{Vacant channel}
		\State Switch to current channel and enters into the  \textbf{non-Master} mode.
		
		\ElsIf  {Collisions on \textit{$CS$} and \textit{$CT$} slots}
		\State Switch to current channel and enters into the  \textbf{non-Master} mode.
		\EndIf
		
		\EndFor
	\end{algorithmic}
\end{algorithm}

\begin{algorithm}[!b]
	\caption*{\textbf{Subroutine 4:} Non-Master Mode}
	\begin{algorithmic}[1]
		\State Input: $\pi_{n,r},K, MB  \text{ (Current channel of master)}$
		\For {$SB$=1,2,....,K-1}
		\If {Collision observed in \textit{$CT$ slot}}		
		\State  Decide if like to switch with the channel MB.
		\Else 
		\State Continue on the current channel.
		\EndIf
		\If {Switch request accepted}
		\State Transmit in the $\textit{CS}$ slot and shift to channel MB.
		\State Update $\pi_{n,r}=MB$.
		\Else 
		\State Do not transmit in the $\textit{CS}$ slot.
		\EndIf
		\EndFor
	\end{algorithmic}
\end{algorithm}

\subsection{Pseudo Code}
The pseudo-code of the SMCS phase is summarized in Subroutines $2,4,5$. $\pi_{n,r}$ indicates the reserved channel of the user $n$ and it is the channel on which the user had recently locked. For example, $\pi_{n,r}=\pi_{T_{rh}}(n)$ when user enters in the SMCS phase. After entering into the SMCS phase, the user can either be in master or non-master mode based on the index of the reserved channel (lines $4-8$: Subroutine $2$). In the master mode (Subroutine $3$), master identifies the index of the SB. As mentioned before, in the first SB of each MB, the master transmits on its reserved channel (line $2$: Subroutine $3$). In the rest of the SB blocks, master selects the channel based on the preference list obtained using the \textcolor{black}{UCB} algorithm (line $4$: Subroutine $3$). As discussed before, master moves to non-master mode in one of the three scenarios: 1) No collision in CT slot (lines $5-6$: Subroutine $3$), 2) 1) Collisions in CT and subsequent CS slots (lines $7-8$: Subroutine $3$), and 3) End of the MB block (lines $2$: Subroutine $3$).

The pseudo-code for channel selection in non-master mode is given in Subroutine $4$. If the non-master faces collision in the CT slot, it checks whether the reserved channel of the master (i.e.,the channel with index MB) is better than her current channel. If yes, non-master accepts the switch request by transmitting in the subsequent CS slot, updates and moves to the new reserved channel (lines $8-10$: Subroutine $4$). Otherwise, she remains silent in the CS slot indicating switch reject (line $12$: Subroutine $4$). If there is no collision in the CT slot, non-master transmits on the reserved channel in the CS slot (line $6$: Subroutine $4$).

\subsection{Analysis}
We analyze the performance of the $dSOC\_SN$ algorithm and show that it leads to a stable orthogonal configuration (SOC). Our main result is the following theorem.

\begin{thm}
\label{thm:SOC_Static}
Consider a network with $K$ channels and $N$ users with channel rewards characterized by $\{\mu_{n,k}\}$ for all $n \in [N]$ and $k \in [K]$. For any $\delta>0$, set $T_{rh}$ as in Eq.~\ref{eq:trh}. Then, there exists $T(\delta)$ such that for all $t \geq T_{rh} (\delta)+ T(\delta)$, the probability of the network being in an SOC is at least $1-2\delta$.  
\end{thm}
We prove the result using the following lemma. Its proof is given in the appendix. 
\begin{lemma}
	\label{lma:RanHop_Static}
	Let $\delta \in (0,1)$. If RH sub-phase is run for  $ T_{rh}(\delta) $ number of time slots, then all the users will orthogonalize with probability at least $1-\delta$	where 
	\begin{equation}
	\label{eq:trh}
	  T_{rh}(\delta):=\left \lceil \frac{\log(\delta/K)}{ \log\left(1-1/4K\right)} \right \rceil
	\end{equation}
\end{lemma}  
The lemma guarantees that the users are orthogonalized with probability at least $1-\delta$ at the end of the RH sub-phase. We next prove the theorem conditioned upon this event. The proof is an adaptation of the proof of Thm $1$ in \cite{lax} to our specific block structure. 

{\em Outline of Proof of Thm.  \ref{thm:SOC_Static}:}  Note that the block structure is designed such that even if user's request to swap on a channel is rejected, she still gets to observe a reward/throughput sample from that channel (in CS slot). Thus the master gets to explore the channels in a master block as if she is the only user in the network. Also, each user gets to observe many samples of each channel due to the inherent exploration component in the UCB Index given in Eq.~\ref{eqn:UCBInd}. 
\noindent
Then, following the same arguments as in the proof of Lemma $1$ in \cite{lax}, if all channels are sampled at least $s_{\min}$ number of times \textcolor{black}{by time $t$}, then all the users will have a correct ranking of the channels (with respect to the true means) with probability at least $1-2t^{-4}$. Thus, any channel switch thereafter results in both users moving to their better channels leading to decrease in network potential. The value of $s_{\min}$ is given by
\begin{equation}
\label{eqn:MinSamp}
s_{\min}:=\frac{8 \log t}{\Delta_{\min}^2} 
\vspace{-0.2cm}
\end{equation}
where \vspace{-0.1cm}
\[\Delta_{\min}=\min_{n \in [N]} \Delta_n \mbox{ and } \Delta_n= \min_{i,j \in [N], i\neq j} |\mu_{n,i}- \mu_{n,j}|\]

However, for the specific block structures designed for coordination, each user may not get an informative reward sample in each time slot. 
A master in a MB will get at least $2 + K-1$ samples, while a non-master will get at least $2K-2$ samples. Hence over an OHS block consisting of $2K^2$ slots, each user will get at least $(K-1)(2K-2)+ K+1$ informative samples. Also, the minimum number of samples given in Eq.~(\ref{eqn:MinSamp}) should be satisfied for all channels. 
Hence, the condition on the smallest $t$ such that Eq.~(\ref{eqn:MinSamp}) holds across all channels is given by
\[t\geq K\frac{2K^2}{(K-1)(2K-2)+ K+1}s_{\min}\geq \frac{16K}{\Delta_{\min}^2}\log t.\]
Again using \cite{lax}[Lemma 1], we can show that the smallest $t$, denoted $t_m$, satisfying Eq.~(\ref{eqn:MinSamp}) is finite and satisfies
\[t_m \leq \frac{M-1-\sqrt{(M-1)^2- 4M}}{2}, \mbox{  where } M:=\frac{16K}{\Delta_{\min}^2}.\] 
Thus, for all $t \geq t_m$ every channel switch results in both users switching to better channels and the network potential decreases.

Further, note that the maximum value of the network potential is at most $N(K-1)$. \textcolor{black}{In each OHS each player gets a chance to switch channel by becoming a master. Hence it takes at most $K-1$ OHS slots to reach stable allocation after $t > t_m$. As potential decreases in a channel switch happens provided the users have correct ranking of the channels in that round. Hence the probability that the network is in SOC within $\tau=2K^2(K-1)$ (each user got $K-1$ attempts to switch) slots after initial $t_m$ slots is at least} 
\[P_{soc}= (1-2t_m^{-4})^{N(K-1)}.\] c
For any $t$, let $S_t=1$ and $S_t=0$ denote the events that network is in SOC and not in SOC, respectively. We have
\[\Pr\{S_{\tau+t_
m}=1| S_{t_m} =0  \} \geq P_{soc},\]
Then, for any $T> \tau + t_m$
\[\Pr\{S_T=0 | S_{t_m}=0  \} < (1-P_{soc})^{\frac{T-t_m}{\tau}}.\]
Setting $(1-P_{soc})^{\frac{T-t_m}{\tau}}\leq \delta$ and solving we get 
\begin{equation}
\label{eqn:T_delta} T(\delta):= t_m+ \tau \log \left(\frac{\delta}{1-P_{soc}}\right)=\textcolor{black}{O(1/\Delta_{\min}^2 + K^3\log(1/\delta)).}
\end{equation}
Then, for all $T\geq T(\delta)$ the network will be in a SOC with probability at least $(1-\delta)$. Taking into account the initial $T_{rh}(\delta)$ rounds of random hopping in which orthogonalization happens with probability at least $(1-\delta)$, we conclude that for all $t > T_{rh}(\delta) + T(\delta)$, the network will in SOC with probability at least $1-2
\delta$.  This completes the proof. \hfill\IEEEQED

\section{Dynamic Network: Algorithm and Analysis}\label{sec:DN}
In this section, we consider dynamic networks where users can enter or leave the network anytime without prior agreement. Very few algorithms can be extended for dynamic ad-hoc networks and they assume global  synchronization which means new users have complete knowledge about the status of the network. For example, in \cite{MC,wcnc,rjain}, algorithms exploit the full knowledge of network state and restarts at regular intervals to account for the dynamic users. However, requiring complete knowledge of the network state is restrictive as non-active users need to continuously sense the network without utilizing the energy efficient sleep mode.  We remove such restriction by allowing users to identify the parameters of the block on their own. Our block structure is designed in such a way that the new users can figure out the current state of the network themselves within a few rounds. 

\subsection{dSOC\_DN Algorithm}
When a new user enters into the network, she does not have any knowledge about the current MB and the slot type (CT/CS). Her first task then is to synchronize with the network to identify parameters such as index of the MB and know which slot is CT and CS. In an ad-hoc network where there is neither a central controller nor control channel between users, synchronization is a difficult task unless existing users help or guide the new users. To achieve synchronization among the users without the need of global clock and horizon synchronizations, we develop the dSOC\_DN algorithm with appropriate modifications to the dSOC\_SN algorithm. The pseudo-code of the dSOC\_DN algorithm is given in Algorithm $2$. It consists of two phases: 1) Synchronization phase, and 2) SMCS phase. The synchronization phase enables users to identify block parameters such as $M_{ind}$ and its reserved channel, $\pi_{n,r}$ while the SMCS phase is identical to that in the dSOC\_SN algorithm with two modifications discussed below to aid the synchronization.


\begin{algorithm}[!h]
	\caption{dSOC\_DN Algorithm} \label{Algo2}
	\begin{algorithmic}
		\State	Input: {$K$}
		\State	$M_{ind}, \pi_{n,r}= SP(K)$ 
		\State	 SMCS $(\pi_{n,r},K,1)$ 
	\end{algorithmic}
\end{algorithm}

\subsection{Modified SMCS Phase}
Recall that in the dSOC\_SN algorithm all active users transmit on their reserved channels in the first sub-block of each MB. In the modified SMCS phase, only the master is allowed to transmit on the reserved channel while other users remain silent in the first sub-block of each MB. The first sub-block of each MB is referred to as synchronization sub-block (SSB) as it will help new users to synchronize in the network. For illustration, if a new user observes that no transmission happen for two consecutive time slots on an occupied channel, then she 
knows that the first slot where no transmission happened is the CT slot. Thereafter, slots alternate between CS and CT. The second modification prohibits the users to leave the network when they are in the non-master mode. Specifically, when a user has to leave she will do so only at the start of the MB where she is supposed to be the master. Without such restriction, the new user will not have sufficient information to identify the block parameters using SSB and she will have to frequently switch to other channels whenever the existing user leaves the network at arbitrary times. 

\textcolor{black}{Note that the leaving user has to delay its departure by at most one OHS duration, i.e.,  is $2K^2$ slots, which is a vanishing portion of the horizon size and hence this assumption is not overly restrictive. This restriction applies only to the leaving users and the new users can enter any time. 
Existing algorithms also put such restriction on entering and leaving users. For instance, \cite{MC} does not allow the users to enter and leave during the learning period which is significantly large of the order $max(16K/\Delta^2, 50K^2)$. Similarly, \cite{wcnc} does not allow players to leave during learning period of duration approximately $max(2K/\Delta^2)$.}

\subsection{Synchronization Phase}
A new user starts with the synchronization phase after entering into the network. In this phase, the new user randomly selects a channel in each slot until it finds a channel that is occupied by another user. Once the new user finds an occupied channel, it stays on that to find the network status. Such channel is used by the new user to find network state and it is referred to as 'piggyback channel' while the user on the piggyback channel is referred to as \textcolor{black}{ 'piggyback user'}. We say that a new user has entered into the piggyback phase once she finds an occupied channel. In piggyback phase, the user senses the same channel continuously till it observes no transmission for two consecutive time slots immediately followed by at least one transmission. After identifying such time slots (i.e., SSB), a user can easily differentiate between CT/CS time slots. However, a new user cannot know the index of the MB block, $M_{ind}$, without which it cannot enter into the SMCS phase. 

In order to find $M_{ind}$, a new user has to sense the piggyback channel until one of the two events happen: 1)~Piggyback user becomes master, or 2)~Piggyback user leaves the network. When a piggyback user becomes master, a new user can sense transmissions instead of silent SSB. Similarly, when a piggyback user leaves the network, a new user will sense silent slots for at least $2K$ time slots. In each case, $M_{ind}$ is same as the index of the piggyback channel. Note that both these events can happen only once in the OHS block and since the duration of OHS block is $2K^2$ time slots, a new user must sense the piggyback channel for $2K^2$ time slots in the piggyback phase. After that, the new user is guaranteed to have estimated $M_{ind}$ and can enter into the SMCS phase.

After synchronization and before entering into the SMCS phase, the new user needs to have its own reserved channel. It is identified by sequentially sensing the channels until she finds a vacant channel which is not occupied by any of the active users. Note that new user cannot take the reserved channel of the current master and this can be easily avoided as she has complete knowledge of block parameters.

The pseudo code of the proposed synchronization phase is given in Subroutine 5. When a new user enters into the network, she selects the channel uniformly randomly (line 5) and senses it. The user enters into the piggyback phase if the channel is sensed as occupied (line 6). When the user senses the channel as vacant for two consecutive time slots, the user is said to be synchronized ($Sync=1$) (line 9). Thereafter, user senses the channel for at most $2K^2$ time slots to identify the $M_{ind}$ (line 12) and enters into the SMCS phase after identifying the reserved channel (line 13).


\begin{algorithm}[!h]
	\caption*{\textbf{Subroutine 5:} Synchronization Phase for New User}
	\begin{algorithmic}[1]
		\State Input: $K$  
		\State Set $Piggyback=0$ and $Sync=0$.
		\While{$Sync==0$}
		\While {$Piggyback==0$}
		\State Sense randomly chosen channel, $\pi_n(t)$$\sim$$U(1,..,K)$. 
		\State Enter into Piggyback phase,  $Piggyback=1$, if channel is occupied
		\EndWhile
		\State Sense the same channel, $\pi_n(t) =\pi_n(t-1)$.
		\State Synchronization done, $Sync=1$, when channel is sensed as vacant for two consecutive time slots followed by at lease one transmission.
		\EndWhile
		\State Sense the same channel for $2K^2$ time slots.
		\State Identify the index of the MB. 
		\State Identify reserved channel and enter into Subroutine 2: SMCS phase.
	\end{algorithmic}
\end{algorithm}

\begin{figure}[!h]
	\centering
	
	\includegraphics[scale=0.275]{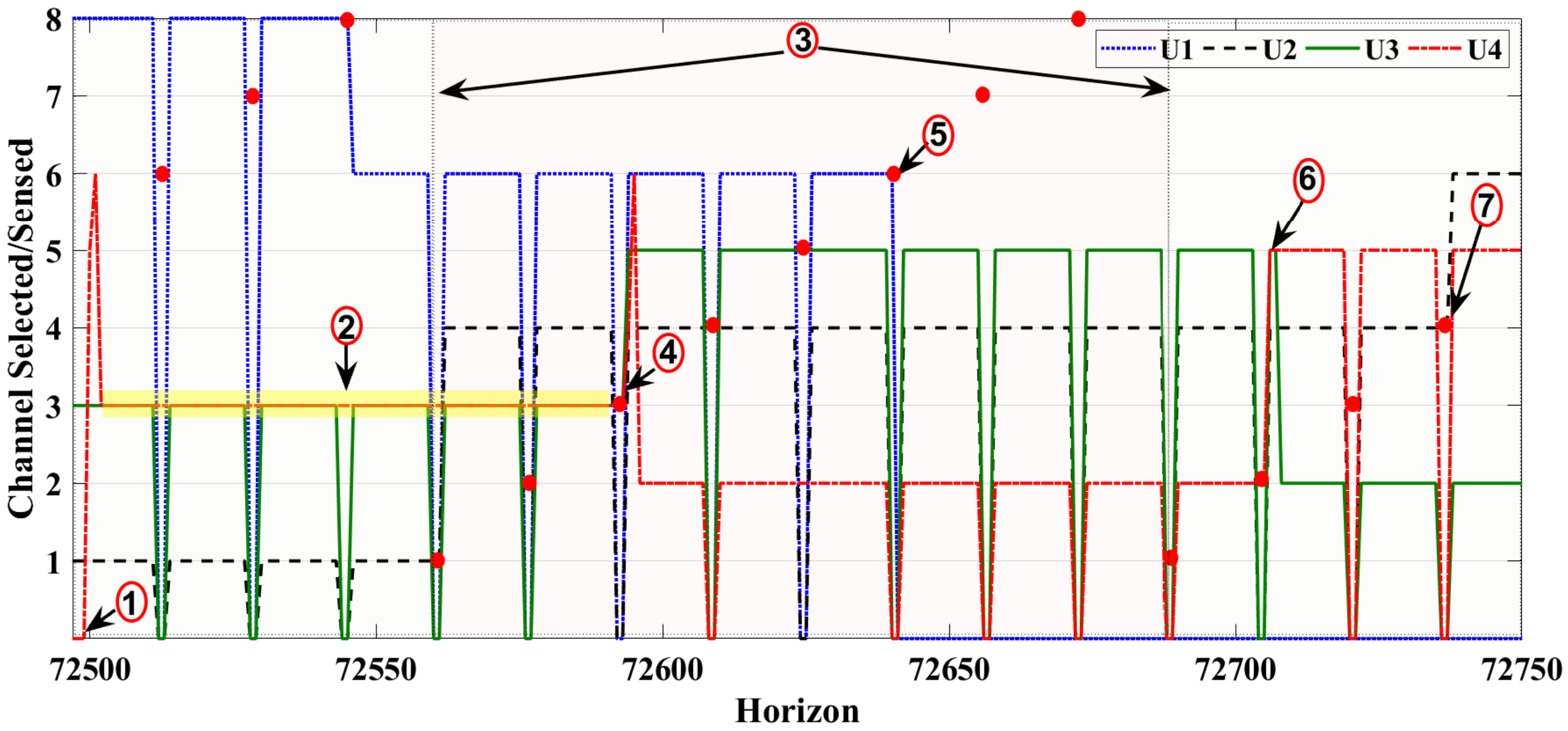}
	\caption{\footnotesize{Channel selection/sensing by different users in dSOC\_DN algorithm with entry of new user as well as exit of the existing users.}}
	\label{chsel}
\end{figure}

Next, we demonstrate the switching from synchronization to SMCS phase using a suitable example. For illustration, we highlight the channel selection of various users during certain interval of the horizon, say  $t = 72490$ to $t=72750$. As shown in Fig.~\ref{chsel}, $x$-axis represents time, $y$-axis represents the channel index and a number inside the circle indicates a particular action. The index of the MB is shown using red colored dark circles and OHS regions are indicated with the different colors. For instance, action 3 shows the boundary between OHS blocks. The Fig.~\ref{chsel} begins with fifth MB of OHS block and comprised of one complete OHS block followed by four MBs of the next OHS block. There are three users (U1, U2, and U3) in the beginning and their channel selections are indicated using different lines. Action 1 indicates entry of new user, U4, at $t = 72500$. After entering into the network, U4 selects channel 5, 7 and 3 uniformly random and senses it. Once she senses the channel 3 as occupied, she enters into the piggyback phase. The duration of the piggyback phase is indicated using a yellow shaded region (action 2). As soon as U3 becomes a master, U4 completes its piggyback phase (action 4), identifies the channel 2 as its reserved channel and enters into the SMCS phase. You can also observe the channel swapping or switching between master and non-masters at different instants in Fig.~\ref{chsel}. For example, action 5 indicates exit of U1 when she is master while action 6 indicates the channel switching between U3 and U4. Similarly, in the last MB, U2 switches to the channel vacated by U1. In this way, proposed algorithms allow the network to reach SOC within a finite time after every entry or exit of a user. 

\subsection{Analysis}

Next, we analyze the performance of the $dSOC\_DN$ algorithm and show that it leads to a stable orthogonal configuration in finite time after entry or exit of users. Our main result is the following theorem. 


\begin{thm}
	\label{thm:SOC_Dynamic}
	Consider a network with $K$ channels and $N$ users with channel rewards characterized by $\{\mu_{n,k}\}$ for all $n \in [N]$ and $k \in [K]$. For the network in SOC, if $e$ and $l$ are the number of users enter or leave the network, respectively, then after $T_s^d+T^d(\delta)+T_l^d$ number of time slots from the recent entry or exit event, \textcolor{black}{the network will be back in SOC with probability $1-\delta$ where $\delta \in (0,1)$}.
	
\end{thm}
We prove the result using the following lemmas. Their proofs are given in the appendix. 
\begin{lemma}
	\label{lma:newuser_dynamic}
	In a network consisting of $K$ channels and $N$ users, the new user will need at most $T_{s}^d=K(2K+4)+1$ number of time slots to complete the synchronization phase, identify the reserved channel and enter into the SMCS phase.
\end{lemma}

\begin{lemma}
	\label{lma:userenter_dynamic}
	For a network in SOC and $\delta \in (0,1)$, when new user begins its SMCS phase, the network will be in SOC again with probability $\delta$ after $T_e^d(\delta)$ number of time slots assuming no user enters or leaves the network during this period where
	\begin{equation}
	\label{eq:T_e_nu}
 T_e^d(\delta) = t_m^{nu}+\log \left(\frac{\delta}{1-P_{soc}^{nu}}\right)
	\end{equation}
	where
	\begin{equation*}
	\label{eq:tm_nu}
		t_m^{nu} \leq \frac{M^{nu}-1-\sqrt{(M^{nu}-1)^2- 4M^{nu}}}{2}
	\end{equation*}
	\begin{equation*}
	\label{eq:tm_M_nu}
	M^{nu}:=\frac{16(K-N)}{\Delta_{\min}^2} \quad \text{and} \quad P_{soc}^{nu}= 1-2({t_m^{nu}})^{-4}
	\end{equation*}
%
	
\end{lemma}

\begin{lemma}
	\label{lma:userleave_dynamic}
	For a network in SOC and $\delta \in (0,1)$, when one of the users leaves the network, the network will be in SOC again in at most $T_l^d=2K^2(K-1)$ number of time slots provided that no new user enters or leave the network.
\end{lemma} 

%
{\em Outline of Proof of Thm.  \ref{thm:SOC_Dynamic}:} The time required for the network to be in SOC depends on the duration of three events: 1) Time required for a new user to enter into the SMCS phase, $T_s$ (Lemma 2), 2) Time required for a new user to learn the channel statistics and minimize the network potential, $T^d(\delta)$ (based on Lemma 3)), 3) Time required for the network to reach SOC after an exit of the user, i.e. $2K^2(K-1)$ time slots (Lemma 4). 

To find $T^d(\delta)$, we do following modifications in Lemma 3 and Theorem~\ref{thm:SOC_Static}. We replace $M^{nu}$ with $M^d$ where $M^d:=\frac{16(K-N-l)}{\Delta_{\min}^2}$ since new user has to learn the qualities of $(K-N-l)$ channels where $l$ is the number of users left the network. Similarly, we replace $P_{soc}^{nu}$ with $P_{soc}^d$ such that $P_{soc}= (1-2t_m^{-4})^{e(K-1)}$ as there are $e$ new users and maximum possible decrease in potential can be $e(K-1)$. Here, $P_{soc}^{nu}$ indicates the probability that the network is in SOC within $\tau^d=2Ke(K-1)$ time slots after initial $t_m^d$ time slots from the slot $e$-th user enters into the network. Based on these modifications, we get

\begin{equation}
\label{eqn:T_delta_d} T^d(\delta):= t_m^d+ \tau^d \log \left(\frac{\delta}{1-P_{soc}^d}\right)
\end{equation}

Then, in $T_s^d+T^d(\delta)+T_l^d$ time slots after recent entry or exit, the network will be in a SOC with probability at least $(1-\delta)$.  This completes the proof. \hfill\IEEEQED

\subsection{More Users than Channels}\label{NgeK}
When the number of users is more than the number of channels, i.e., $N>K$, there can be two possible options in ad-hoc networks: 1) Allow all users to enter into the network using virtual channels, and 2) Restrict some users from entering the network (or in dynamic networks, users can attempt to enter into the network after certain intervals). Existing algorithms such as \cite{MC,MEGA,emilie,gai2,lax,leshem,lugosi,prand} fail when $N>K$. Virtual channels are used in \cite{quek} which requires a central controller to include ($N-K$) virtual channels for collision-free sequential hopping. Furthermore, as shown in \cite{wcnc, wiopt}, it is not an efficient algorithm when $N<K$. 

The dSOC\_SN algorithm handles $N>K$ scenario by offering the second option in the RH phase. When a user gets locked on the channel in the RH phase, it enters into the SMCS phase after $T_{rh}$ time slots, i.e. at the end of RH phase, and those who do not lock on a channel can leave the network in at most $T_{rh}$ time slots. By the end of the RH phase, ($N-K$) users will experience continuous collisions within the RH phase which is an indication that all the channels are utilized and hence they can leave the network. 

In dynamic networks, users can re-enter the network after a certain interval which depends on the rate at which users enter or leave the network. In each case, a new user needs to complete the synchronization phase and identify the reserved channel before entering into the SMCS phase. The new user cannot find the reserved channel when $N\geq K$. Thus, in $T_r$ (See Lemma 2) time slots after identifying the $M_{ind}$, new user can realize the unavailability of the reserved channel and leaves the network. As long as users remain active in the network for a short duration, new users can enter the network whenever channels are available. In this way, our algorithms handle the case of $N>K$ without compromising on the stability of the network.

\section{Experimental Results}
\label{sec:exp}

To demonstrate the effectiveness of the proposed algorithms,  we present the simulation results for comparison with respect to parameters such as: 1) Network potential, 2) Average and total reward/throughput, 3) Number of channel switching, and 4) Number of collisions. Initially, we consider $K=10$ and $N=\{5,10\}$. Each numerical result presented here is obtained after averaging over 100 independent experiments and the horizon size is 100000 time slots. The channel statistics are unknown, heterogeneous and chosen randomly in each experiment.

\subsection{Static Network}

For static networks, the performance of the dSOC\_SN algorithm is compared with the state-of-the-art Coordinated Stable Marriage Multi-Armed Bandit (CSM-MAB) algorithm in \cite{lax}. Note that CSM\_MAB needs wideband sensing receiver consisting of two parallel ASP blocks and computationally intensive digital baseband processing algorithms compared to a single antenna and ASP-based narrowband radio for the proposed algorithms. \textcolor{black}{For benchmarking, we use an allocation obtained by the Hungarian-algorithm (bipartite matching algorithm) with known channel statistics matrix $M=\{\mu_{n,k}\}$. In this allocation, the players are orthogonalized and achieve optimal reward without incurring any collisions.  We also consider heuristic dSOC\_SN, referred to as dSOC\_SN\_H, which differs from dSOC\_SN in two ways: 1) The size of the MB is reduced to $K$ time slots instead of $2K$ time slots, and 2) Master avoids the particular channel for certain interval whenever switch request for that channel gets rejected. If the switch request gets rejected for $i^{th}$ time, the channel will be skipped for subsequent $2^{i}$ OHS blocks and the count, $i$, is reset to zero whenever the master changes her reserved channel. The first modification restricts the number of channels available for switching from $K$ to $K/2$. The second modification minimizes the switching requests to channels which are on the preferred list but have been occupied by other users and they are not willing to switch to the master's reserved channel.} Both these modifications offer a higher number of channel switching opportunities (due to reduced duration of OHS block) resulting in faster orthogonalization and hence, higher reward.

\begin{figure*}[!b]
	\centering
	\subfloat[]{\includegraphics[scale=0.33]{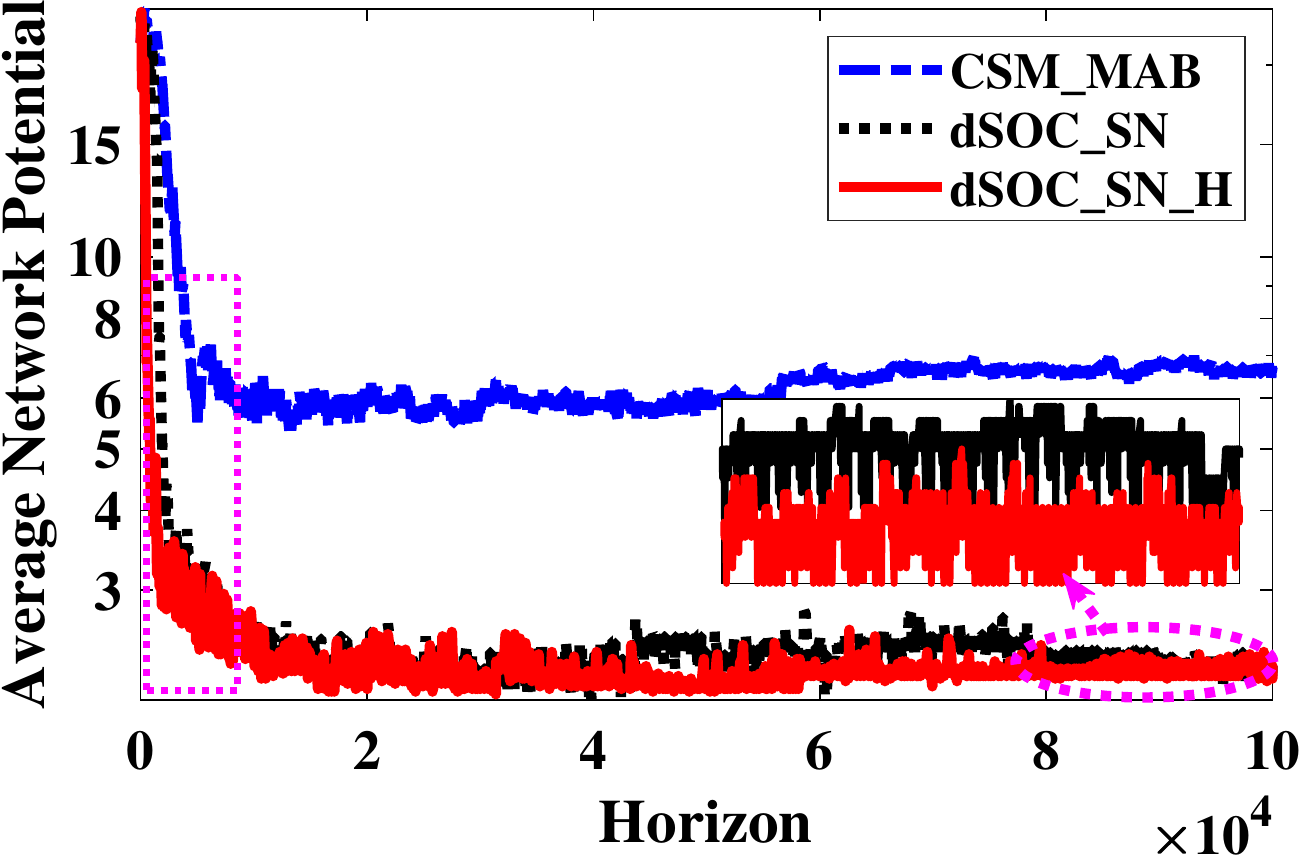}
		\label{N51}}
	\subfloat[]{\includegraphics[scale=0.33]{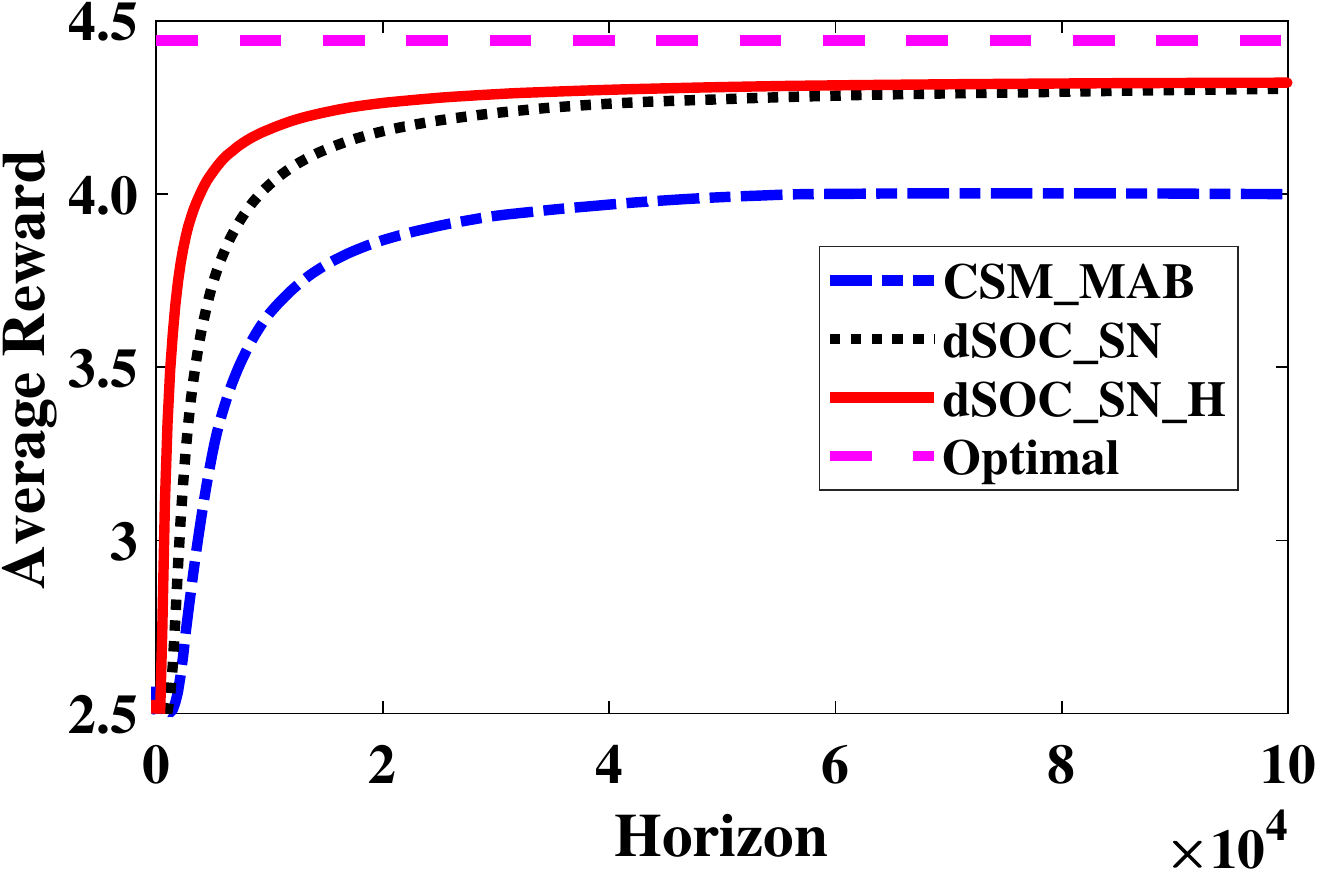}%
		\label{N52}}
	\subfloat[]{\includegraphics[scale=0.33]{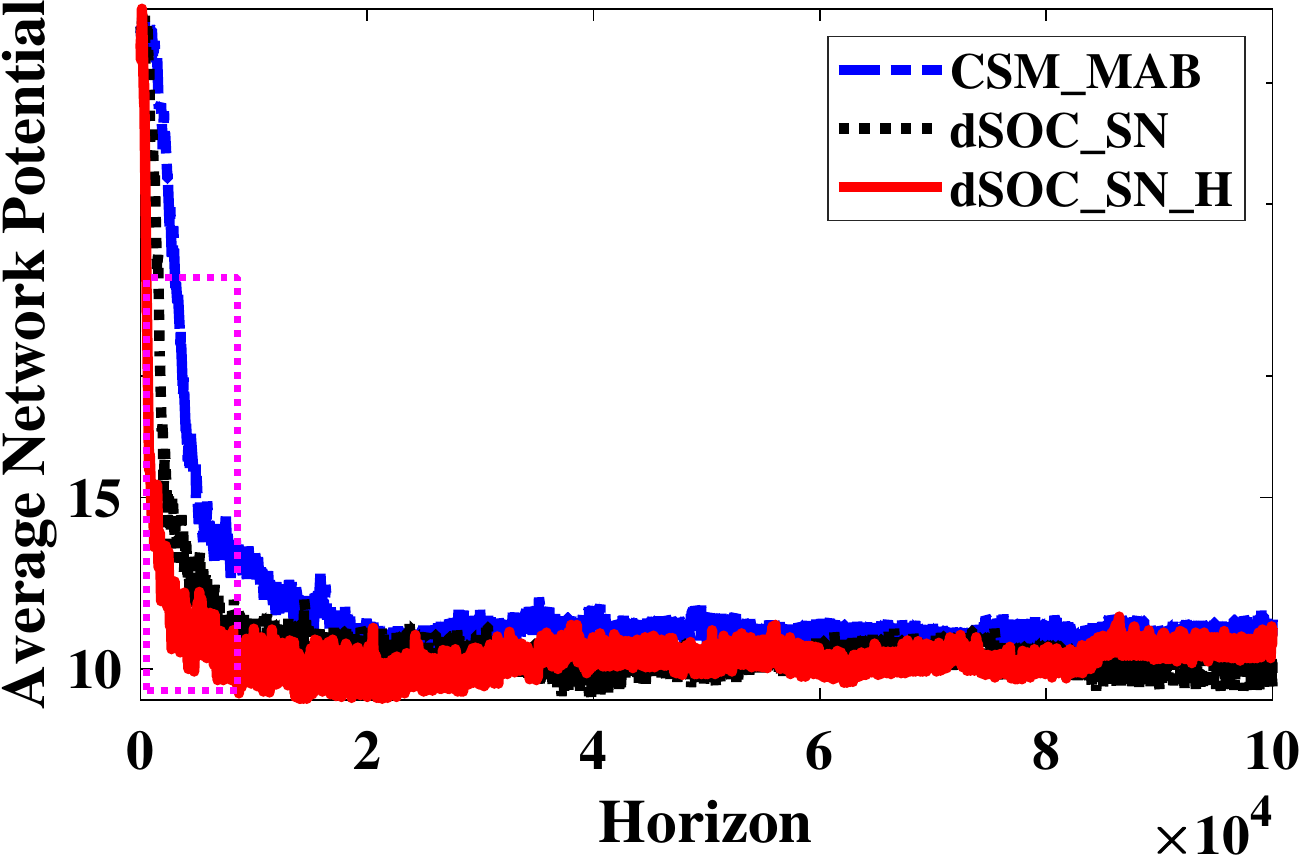}
		\label{N81}}
	\subfloat[]{\includegraphics[scale=0.33]{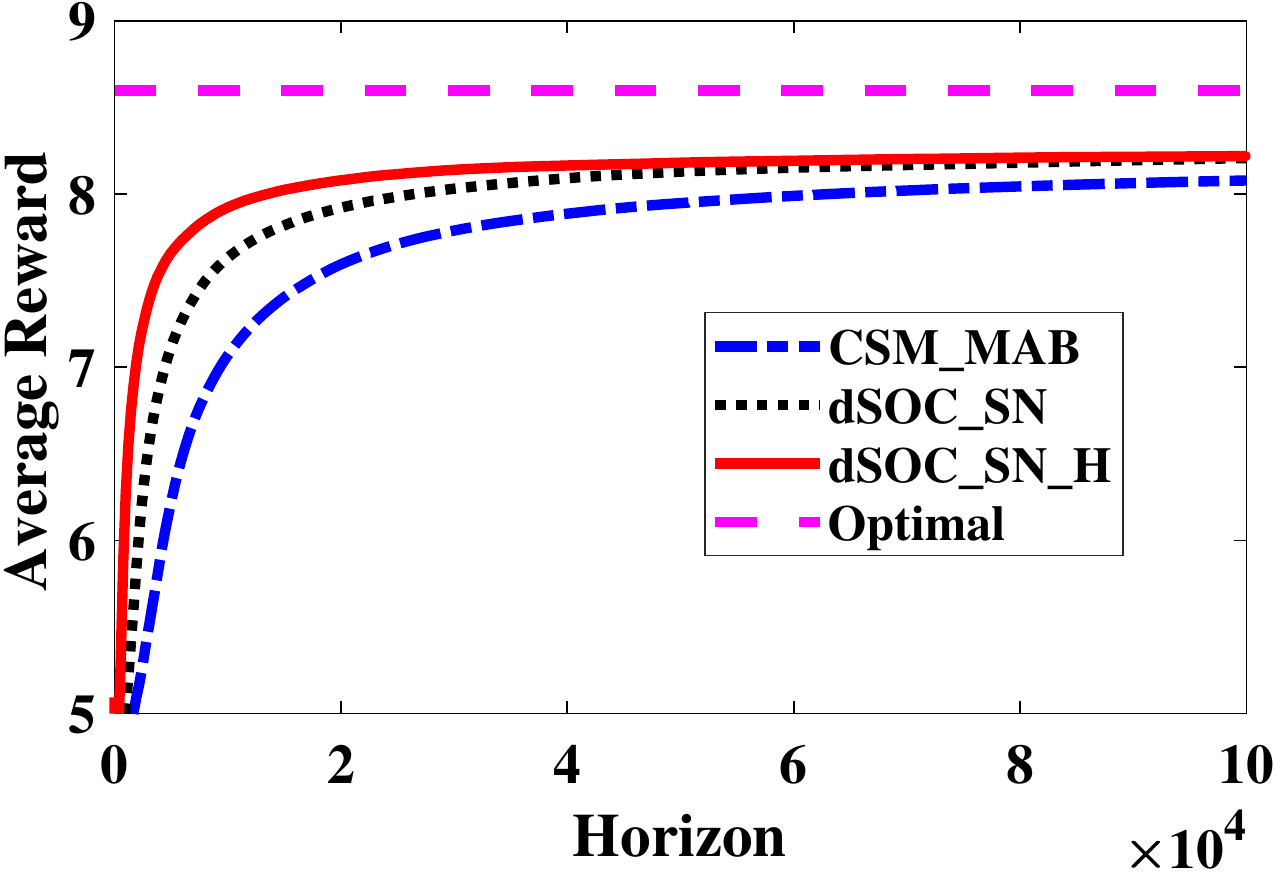}%
		\label{N82}}
	\caption{\footnotesize{Average network potential ((a) $N=5$ and (c) $N=10$) and average cumulative reward of ((b) $N=5$ and (d) $N=10$) users at different instants of the horizon for static ad-hoc network with $K=10$.}.}
	\label{reg2}
	\vspace{-0.25cm}
\end{figure*}

 We first begin with the network potential which is an indication of the time required to reach SOC. The plots showing the variation of network potential with respect to time are shown in Fig.~\ref{reg2} (a) and Fig.~\ref{reg2} (c) for $N=5$ and $N=10$, respectively. The decrease in the potential with time followed by constant potential shows that the corresponding algorithm allows a network to reach SOC. However, proposed algorithms consistently offer a faster decrease in potential and lower average potential than the CSM\_MAB algorithm as highlighted in Fig.~\ref{reg2} (a) and Fig.~\ref{reg2} (c). Next, we consider the average reward at different instants of the horizon. The average reward is the total reward of all user at a given time slot averaged over 100 independent experiments. As shown in Fig.~\ref{reg2} (b) and Fig.~\ref{reg2} (d), proposed algorithms offer a higher reward (and hence, throughput) than the CSM\_MAB algorithm. Significantly higher reward in the beginning ($t<40000$) also indicates early orthogonalization to reach SOC which is a useful characteristic in the short horizon scenario such as dynamic networks. Constant potential and reward plots also indicate that the learning of channel statistics is accurate and proposed algorithms allow users to have sufficient samples of each channel thereby reducing switching to sub-optimal channels. These observations are consistent with Theorem 1.

\begin{figure*}[!h]
	\centering
	\subfloat[]{\includegraphics[scale=0.3]{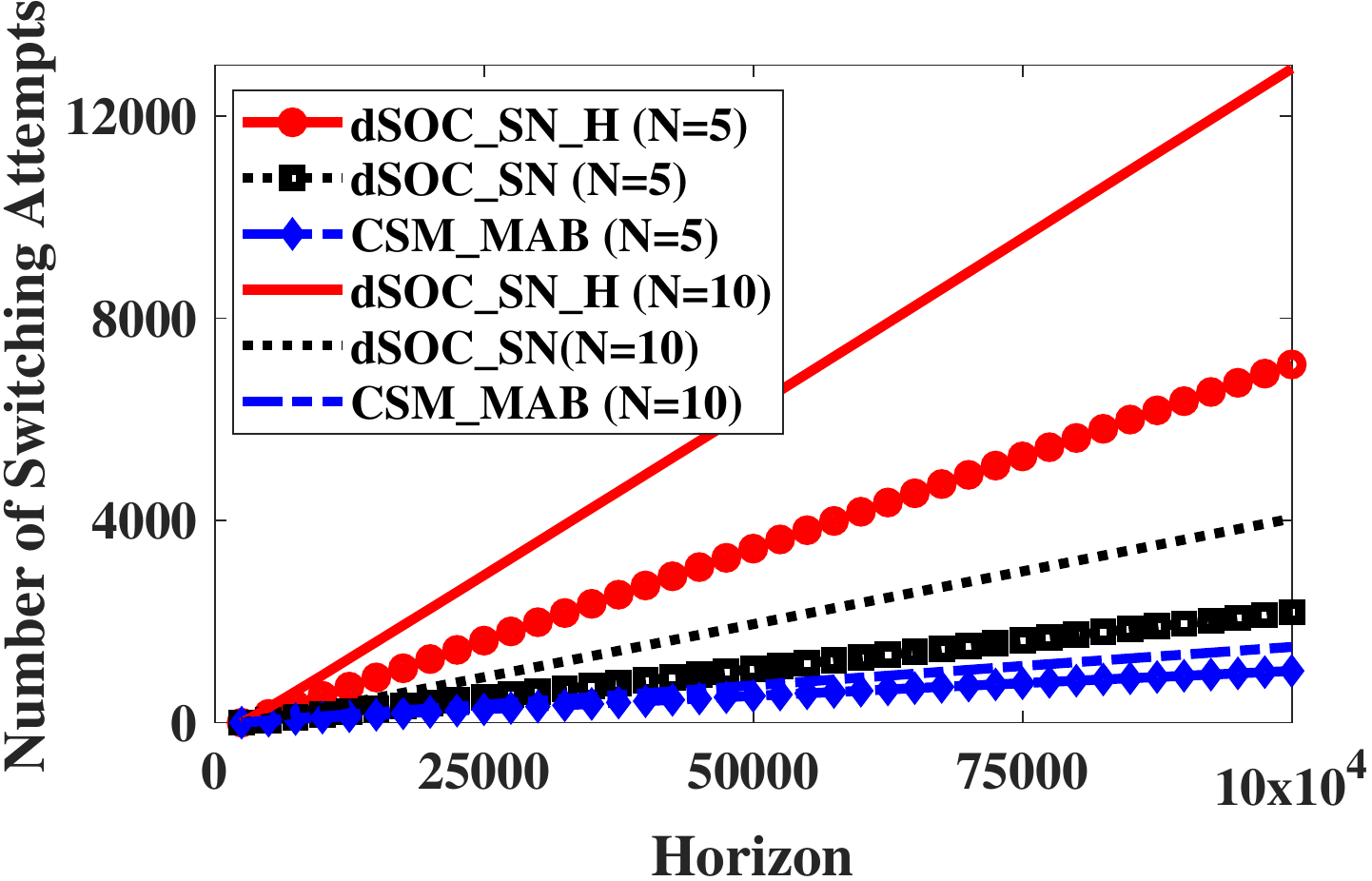}
		\label{CS}}\hspace{0.25cm}
	\subfloat[]{\includegraphics[scale=0.3]{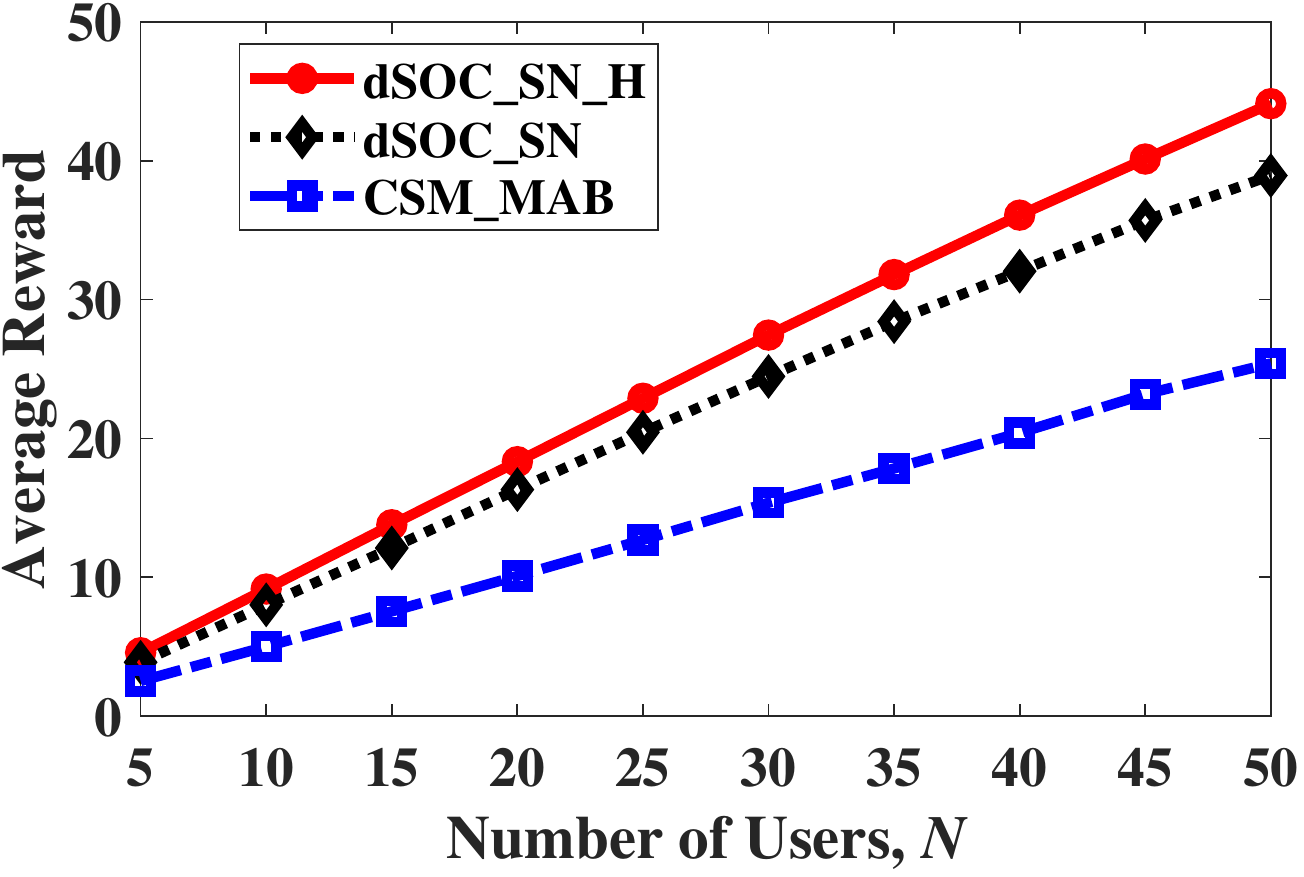}
		\label{lnw_rew}}\hspace{0.25cm}
	\subfloat[]{\includegraphics[scale=0.3]{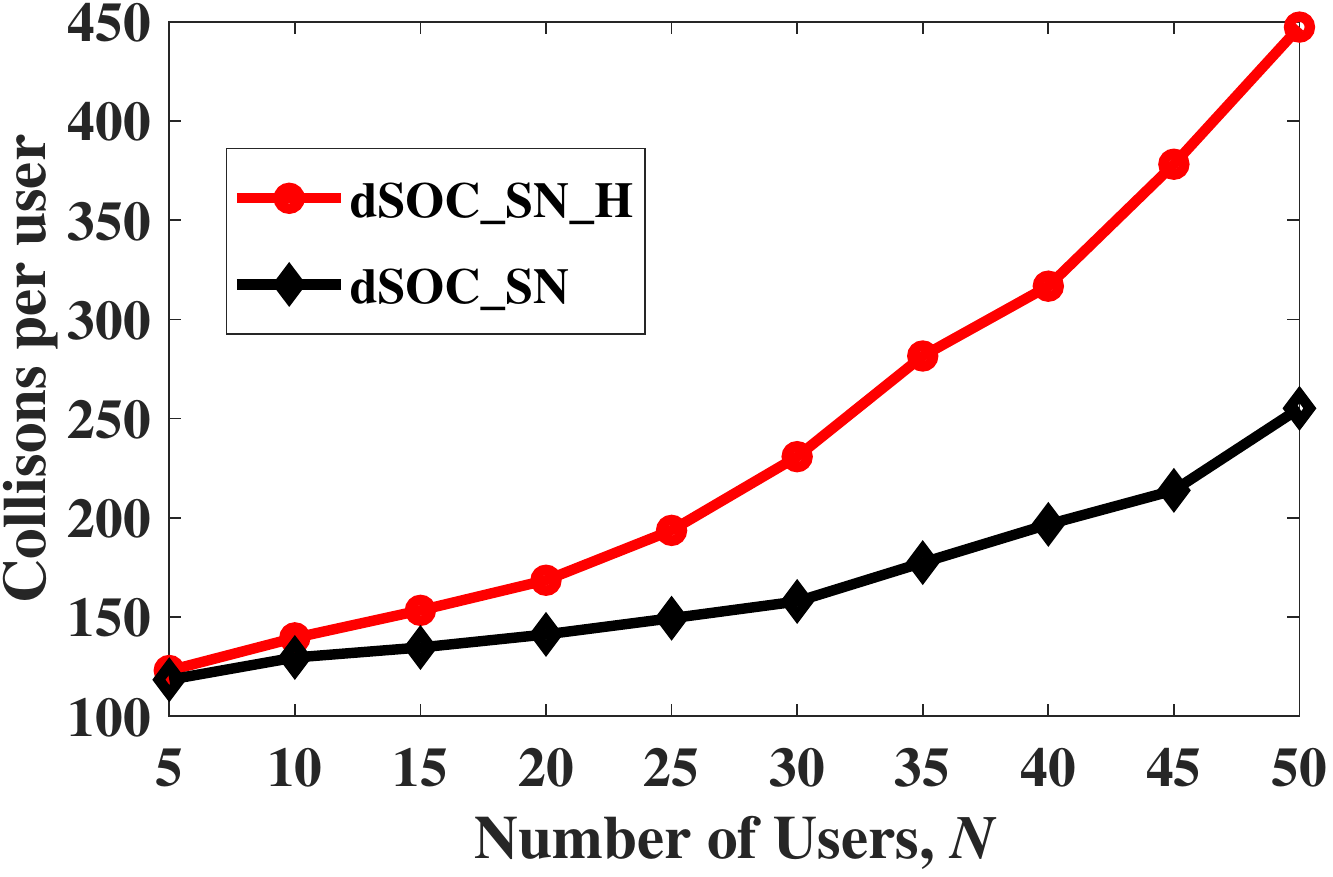}%
		\label{lnw_coll}}
	\vspace{-0.25cm}
	\caption{\footnotesize{a) Comparison of the number of channel switching attempts for static ad-hoc network with $K=10$, b) Average reward of all users, and c) Total number of collisions faced by each user for different values of $N\in\{5, 10,...,50\}$ in the static ad-hoc network with $K=50$.}}
	\label{nk}
	\vspace{-0.25cm}
\end{figure*}

One of the reasons behind the superior performance of the proposed algorithms is that they allow a higher number of channel switching opportunities as demonstrated in Fig.~\ref{CS}. Also, the difference between proposed and CSM\_MAB algorithm increases with the increase in $N$. This is because the proposed algorithms allow each user to become master once in every OHS block while the CSM\_MAB algorithm makes user compete for grabbing the channel switching opportunities. When multiple users compete, no one gets the opportunities leading to poor performance in spite of using complex radios with the wideband sensing capability. Among the dSOC\_SN and dSOC\_SN\_H algorithms, dSOC\_SN\_H offers a higher number of channel switching opportunistic in a given horizon due to reduced duration of the OHS phase.


%

Next, we consider a large size ad-hoc network with $K=50$ channels and $N$ ranging from 5 (sparse network) to 50 (dense network). In Fig.~\ref{lnw_rew}, we compare the average reward of all users for different values of $N$. As expected, the reward increases with the increase in $N$ for all three algorithms. It can be observed that the proposed algorithms offer higher reward than the CSM\_MAB algorithm for all $N$ and the difference increases with the increase in $N$ due to the same reasons discussed above. 

%

Next, we analyze the difference between the dSOC\_SN and dSOC\_SN\_H algorithms based on the number of collisions faced by each user throughput the horizon. Each collision leads to re-transmission of the lost packed leading to wastage of spectrum, time and power. Thus, they should be as small as possible. As shown in Fig.~\ref{lnw_coll}, though dSOC\_SN\_H offers a higher reward, it also leads to a higher number of collisions. The dSOC\_SN algorithm offers approximately half the number of collisions than dSOC\_SN\_H due to longer OHS phase which means fewer switching opportunities as shown in Fig.~\ref{CS}. However, the number of collisions per user is less than 450 for a horizon size of 100000 which corresponds to very small collision probability of 0.005. Thus, proposed algorithms do not incur a large number of collisions even though our signaling scheme for channel switching is based on collision. 
%

\begin{figure}[!h]
	\centering
	\subfloat[]{\includegraphics[scale=0.25]{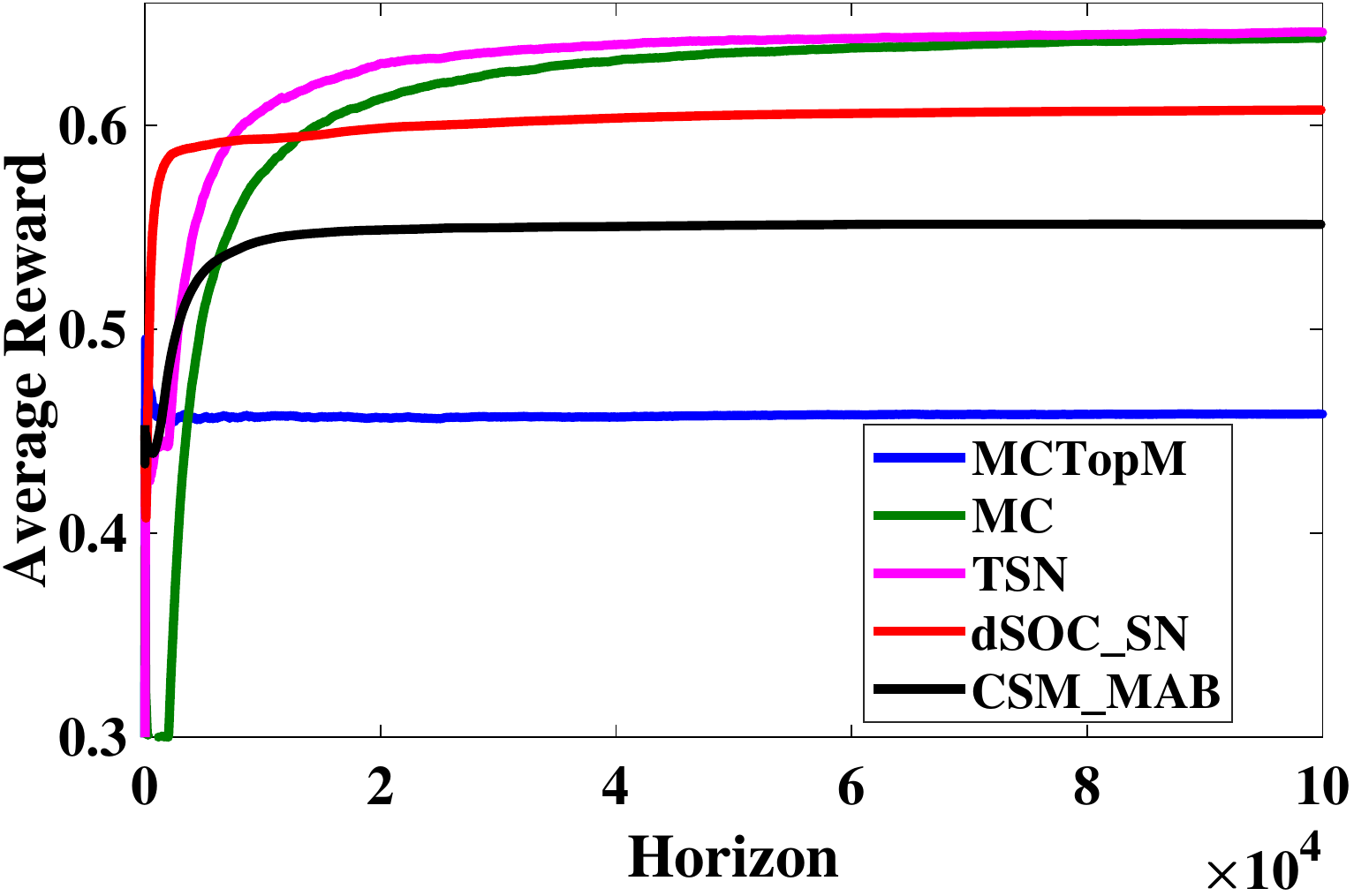}
		\label{fig11}}
	\subfloat[]{\includegraphics[scale=0.25]{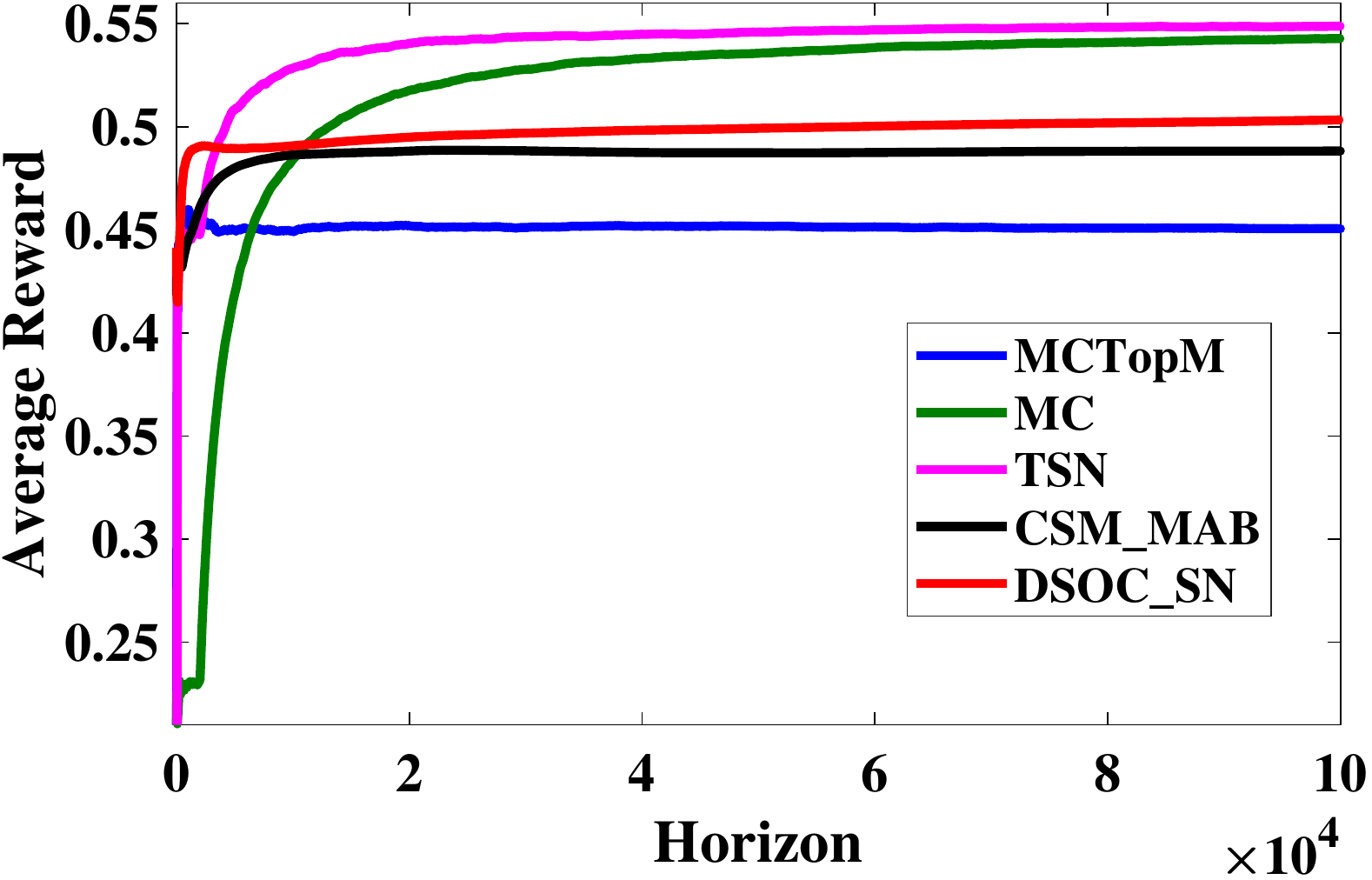}
		\label{fig12}}
		\vspace{-0.2cm}
	\caption{\footnotesize{Average cumulative reward of ((a) $N=4$ and (d) $N=6$) users at different instants of the horizon for static ad-hoc network with $K=8$ homogeneous channels with statistics 0.1, 0.2,..,0.8.}.}
	\label{fig1}
\end{figure}

\textcolor{black}{For homogeneous channel scenario,
      we have considered $K=8$ channels with statistics as $0.1, 0.2,..,0.8$ and $N=\{4,6\}$ users. Corresponding average reward per user plots are shown in Fig.~\ref{fig1}. As seen, the proposed dSOC\_SN algorithm offers better performance than the CSM\_MAB algorithm in [18]. For comparison, we have also compared other well-known algorithms specifically designed for the homogeneous setting (they will not work in the heterogeneous setting). Among them, MCToPM algorithm in [8] needs prior knowledge of $N$ and hence, for a fair comparison, we assume $N=K$. Other algorithms such as MC [12] and TSN [13] algorithms do not need $N$ but they need to a lower bound on the sub-optimality gap $\Delta_{\min}$. 
       As shown in Fig.~\ref{fig1}, the difference between the performance of the dSOC\_SN and other algorithms is very small even though the dSOC\_SN algorithm does not need knowledge of $N$ and $\Delta$ and it works for homogeneous as well as heterogeneous channels.}

\vspace{-0.2cm}
\subsection{Dynamic Network}

For dynamic networks, we consider three scenarios depicting the various combinations of the time interval at which the users enter or leave the network.  We mark the time of entry and exit of the user with a green and black dashed lines, respectively. We set $K$ = 10 and each result shown here is the average of the values obtained over 100 independent experiments with independently chosen channel statistics.

\begin{figure}[!t]
	\centering
	\subfloat[]{\includegraphics[scale=0.275]{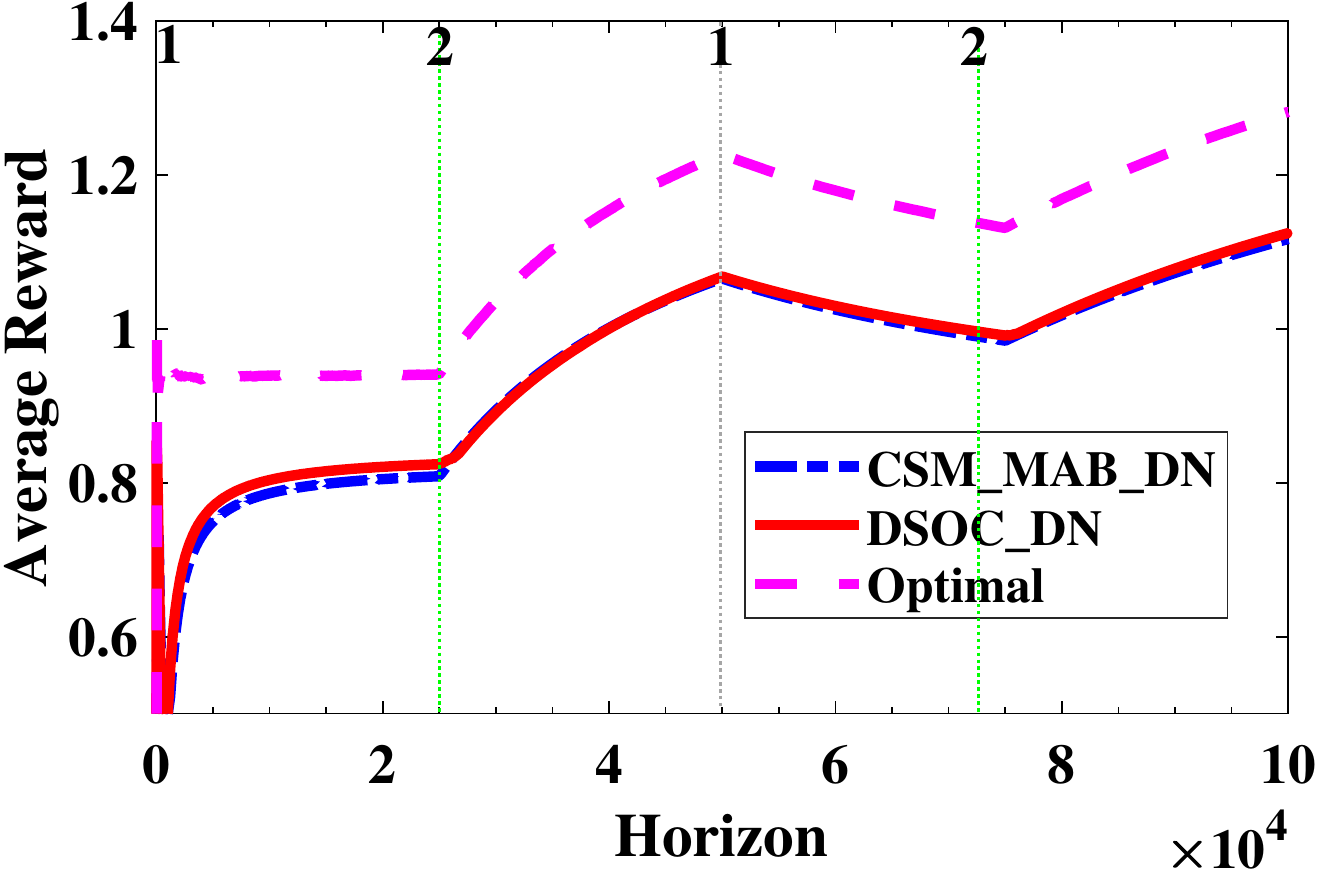}
		\label{DR1}}\hspace{0.05cm}	
	\subfloat[]{\includegraphics[scale=0.275]{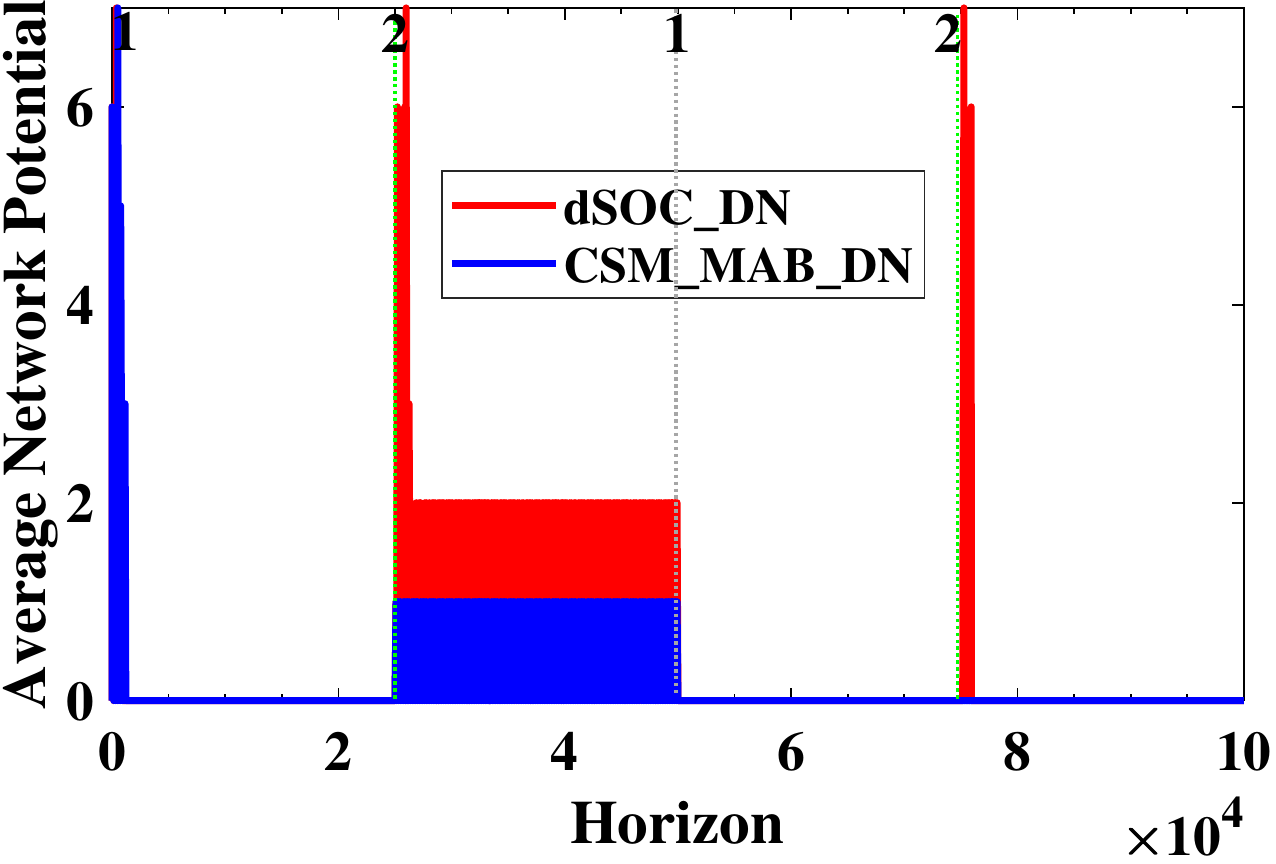}%
		\label{DP1}}\\
	
			\subfloat[]{\includegraphics[scale=0.275]{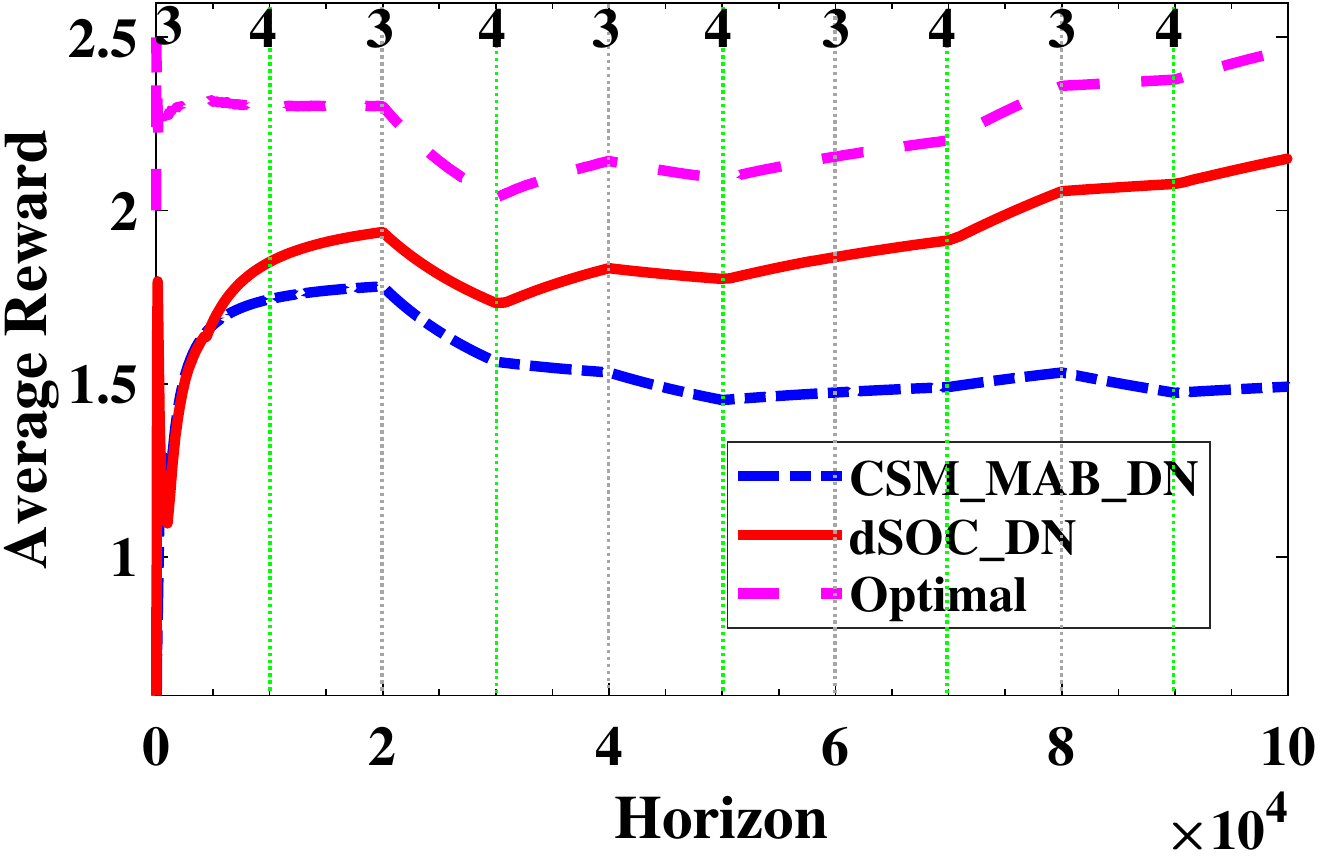}
		\label{DR2}}\hspace{0.05cm}	
		\subfloat[]{\includegraphics[scale=0.275]{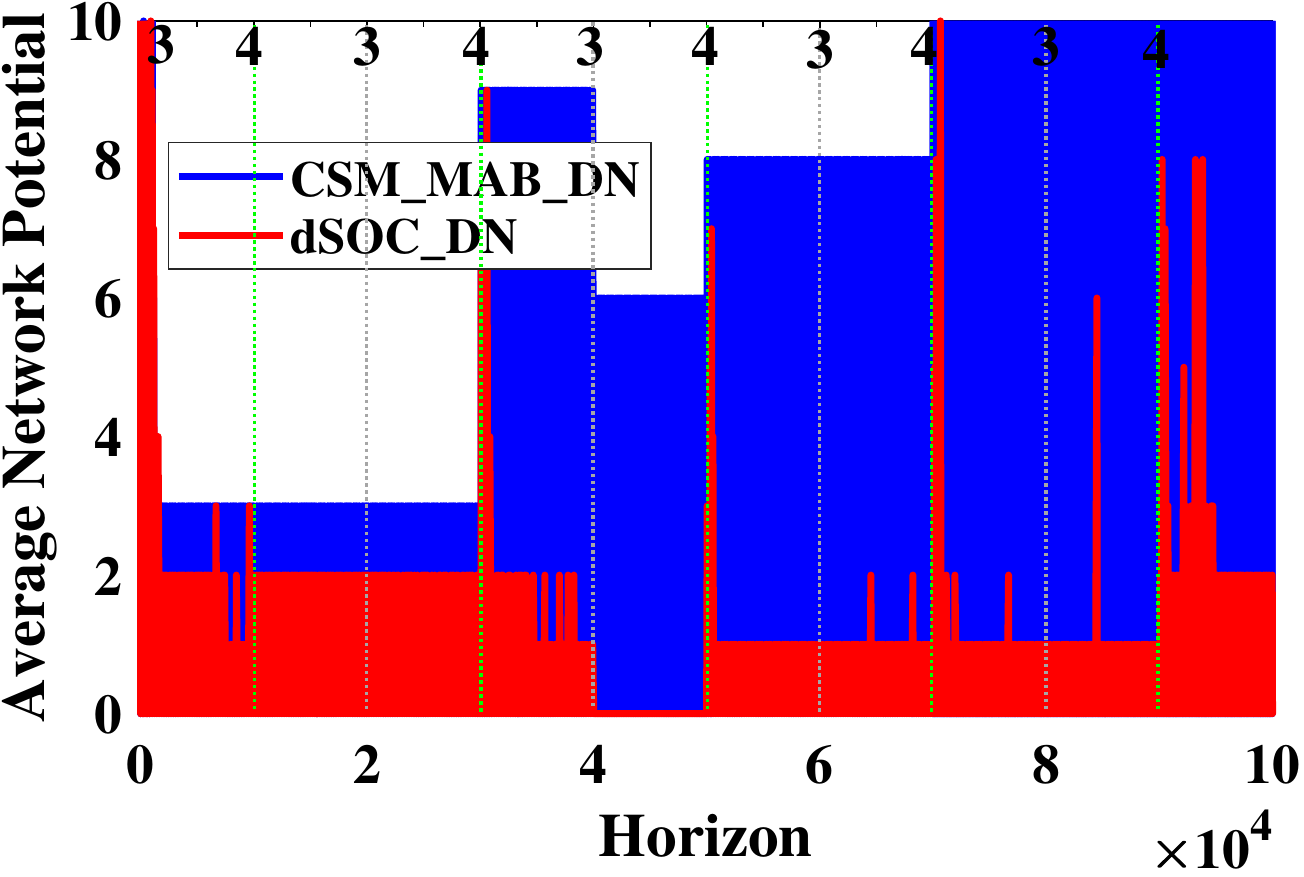}%
		\label{DP2}}\\
	
			\subfloat[]{\includegraphics[scale=0.275]{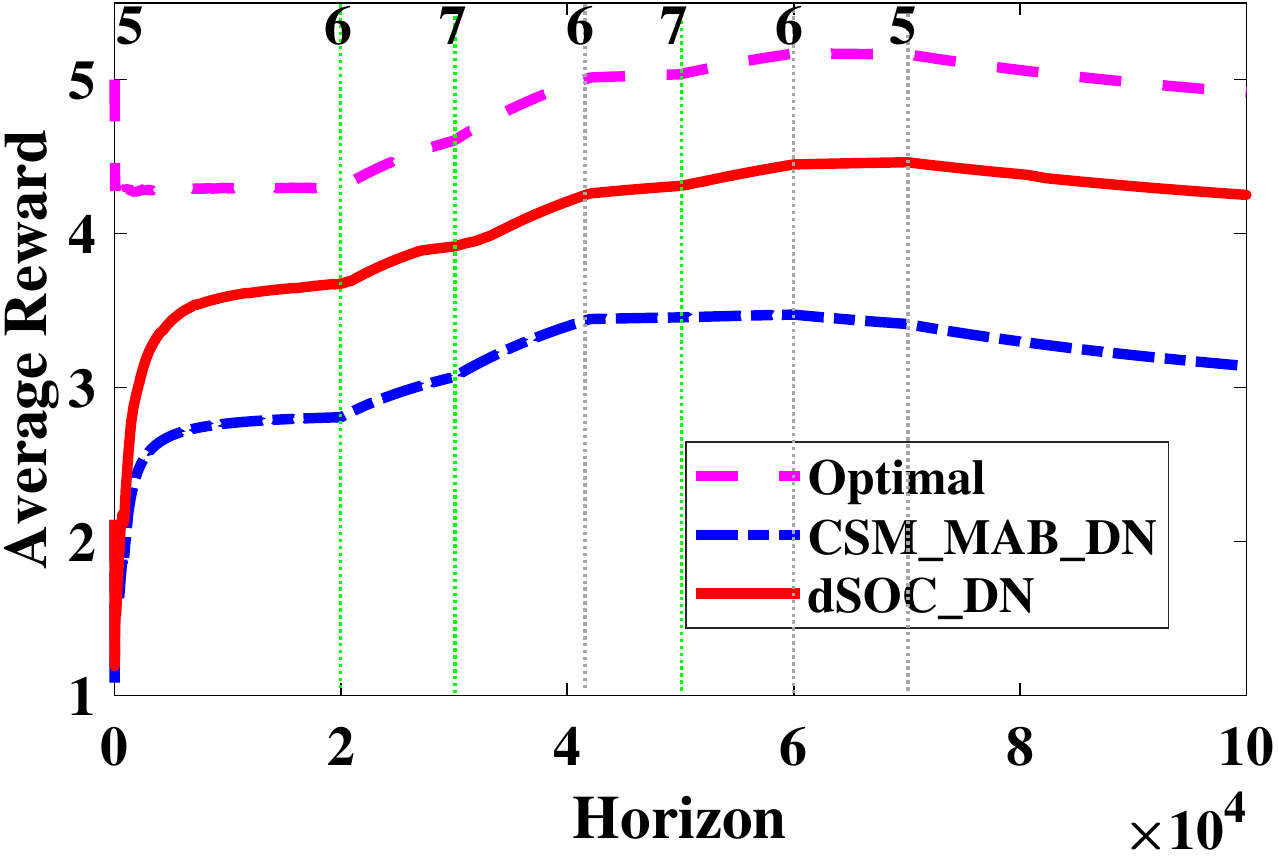}
		\label{DR3}}	\hspace{0.05cm}	
	\subfloat[]{\includegraphics[scale=0.275]{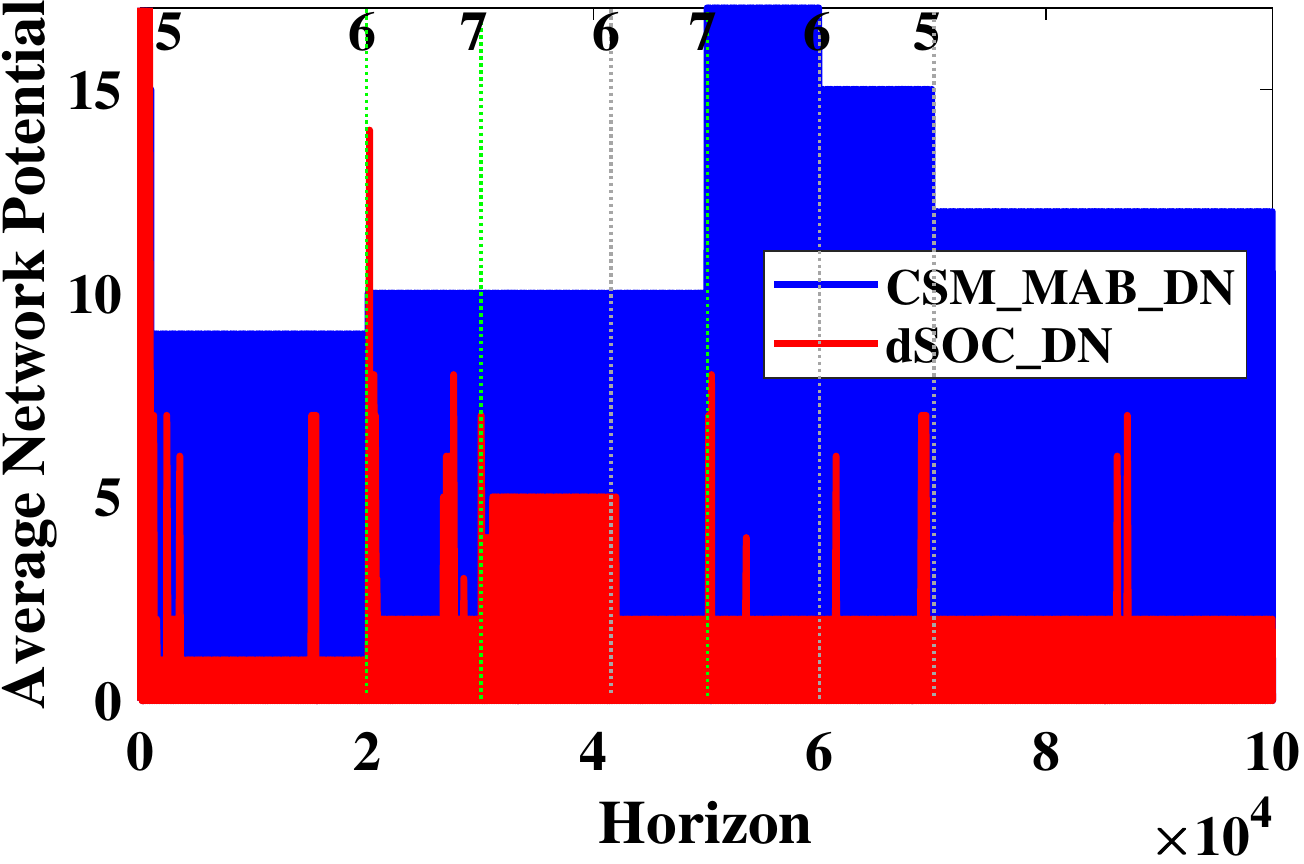}%
		\label{DP3}}
		\vspace{-0.3cm}
	\caption{\footnotesize{Average reward and average network potential comparison between dSOC\_DN and CSM\_MAB\_DN algorithms for dynamic ad-hoc networks with three different scenarios.}}
	\label{rew_dn}
\end{figure}

In the first scenario shown in Fig.~\ref{rew_dn} (a) and (b), there is one user in the network at the beginning. New users enter into the network at $t=25000$ and $t=75000$. Also, a single user leaves the network at $t=50000$ and the leaving user is chosen at random from the set of existing users. As expected, the reward of both algorithms is identical due to the fewer number of users which makes it easy to reach SOC. Note that network potential changes drastically whenever new user enters or leaves the network. This is because, after every entry or exit event, the network may not be in SOC and needs finite time to come back to SOC.

Next, we consider a more challenging scenario with three users in the beginning. Thereafter, at every 10000 time slots, we alternate between user exiting and entering the network with leaving user chosen randomly. It can be observed from Fig.~\ref{rew_dn} (c) and (d) that the proposed algorithm offers significantly higher reward than the CSM\_MAB\_DN algorithm (extended version of the CSM\_MAB algorithm using our synchronization scheme). This is expected as the proposed algorithm has shown to outperform CSM\_MAB for a short-horizon scenario in static networks. Also, the average network potential of the proposed algorithm is lower indicating faster orthogonalization to SOC and a higher number of channel switching opportunities. Note that the network potential increases whenever a new user enters into the network and then decreases with time as network converges to SOC. However, when a user leaves the network, network potential decreases first and it may increase or decrease later depending on the channels vacated by leaving users. For instance, network potential may increase if the channel vacated by the user is sub-optimal for one or more existing users and the users do not have sufficient samples of that channel. In such case, the UCB forces users to explore that channel thereby leading to an increase in the network potential. However, after learning the channel statistics, the network comes back to SOC again. 

We consider the third scenario with more number of users. In the beginning, there are five users and new users enter the network at $T=20000, 30000 \mbox{ and } 50000$ whereas a user leaves the network at $T=42000, 60000 \mbox{ and } 70000$. As shown in Fig.~\ref{rew_dn} (d) and (f), the proposed algorithm offers a higher reward and lower network potential than CSM\_MAB\_DN algorithm. From all three scenarios, we can observe that the difference between the performance of the two algorithm increases with the increase in the number of users, $N$.

\section{Conclusions} \label{conclusion}
In this paper, we presented distributed algorithms to achieve stable orthogonal configuration (SOC) in static as well as dynamic ad-hoc networks. We provided a detailed analysis of the proposed algorithms and validated their performance through simulated experiments for small as well as large size ad-hoc networks. The two novel contributions of the proposed algorithms are: 1) Need of low complexity narrowband radio terminals compared to wideband radios in existing works, and 2) works for dynamic networks where users can enter and leave the system freely. In addition, the proposed algorithms allow new users to synchronize in the network independently without the need for a central controller or continuous sensing when the user is non-active. This feature might be useful for non-active users allowing them to use energy saving sleep mode thereby increasing the lifetime of battery operated radio. 

In the future, we would like to extend the proposed algorithms for guaranteeing optimal orthogonal configuration to achieve higher throughput in addition to SOC. Other interesting scenarios include non-stationary channel statistics, and delayed feedback. For applications such as wireless sensor networks, it is preferable to have terminals without sensing hardware. The design of a distributed algorithm for such terminals is an extremely challenging and open research problem. From the numerical experiments, it appears that if each user's preference list is based on its optimistic value of the current estimates, then the total reward when the network stabilizes is not far from the optimal reward. Another interesting direction is then to characterize the sub-optimality of the total reward from the resulting SoC compared to the optimal reward.

\bibliographystyle{IEEEtran}
\bibliography{biblio}

\appendix
\subsection*{\textbf{Proof of Lemma 1}}
We want to compute $T_{rh}$ such that all users are on non-overlapping channels with high probability within $T_{rh}$. If $P_c$ denotes the collision
probability of a user when all the users are randomly hopping at any time slot $t$, and if none of the other users
are on the non-overlapping channel (worst-case) then the probability that a user will find
a non-overlapping channel within $T_{rh}$ is given by:
\[\sum_{t=1}^{T_{rh}}P_c^{t-1}(1-P_c).\]
We want this probability to be at least $1-\delta/K$ for each user. Hence we set
\begin{eqnarray}
	\lefteqn{\sum_{t=1}^{T_{rh}}P_c^{t-1}(1-P_c) \geq 1- \delta/K}\nonumber\\
	&\iff&1-{P_c}^{T_{rh}} \geq 1- \delta/K 
	\iff T_{rh} \log{P_c} \leq \log\bigg({\frac{\delta}{K}}\bigg)\nonumber \\
	&\iff& T_{rh} \geq \frac{\log\big({\frac{\delta}{K}}\big)}{\log{P_c}}. \label{eqn:TRHBound}
\end{eqnarray}
%
%
\textcolor{black}{
We next give a uniform upper bound on $P_c$. Note that in any round some users may be locked (call them locked users) while others selecting the channel uniformly at random (call them RH users). Fix a time slot $t$ and let $N_{r} \geq 1 $ denote the number of RH users. A user will not incur a collision in round $t$, if it selects one of the channels not used by the locked user (they are $N-N_r$ in number) and further this channel is not selected by any of the RH users. Let $E_1$ denote the event that an (unlocked) user selects one of the channels on which no locked user is present and $E_2$ denote that event that the user sees no collision from an RH user. We have}
\textcolor{black}{\begin{align*}
1-P_c&= \Pr\{ E_1 \& E_2\}=\Pr\{E_1\}\Pr\{E_2|E_1\} \\
	&=  \left ( 1- (N-N_r)/K\right )\Pr\{E_2|E_1\}\\
	&\geq  (1- N/K + N_r/K)\sum_{j=1}^{N-N_r}\frac{1}{N-N_r}(1-1/K)^{N_r-1}\\
	&\geq  (1/K)(1-1/K)^{N_r-1} (\mbox{using $N\leq K$ \& $N_r \geq 1$})\\
	&\geq (1/K)(1-1/K)^{N-1}  \geq 1/4K (\mbox{ for all } K>1).
\end{align*}  
}
Substituting the bound on $P_c$ in Eq.~(\ref{eqn:TRHBound}) and using the union bound we see that within
$ T_{rh}(\delta)$ rounds all the users will orthogonalize with probability at least $1-\delta$. \hfill\IEEEQED

\subsection*{\textbf{Proof of Lemma 2}}
We want to compute $T_{s}^d$ such that new user enters into the SMCS phase and it is the sum of duration of three events: 1)~Time required to enter the piggyback phase ($T_p$), 2)~Time required to identify the CT/CS slots, index of the MB and beginning of the OHS block, ($T_i$) and 3)~Time required to identify the reserved channel, ($T_r$). 

When new user enters into the network, she sequentially senses the channel and enters into the piggyback phase if it is occupied. The worst case corresponds to the network with a single active user and it takes at most $2K$ time slots for new user to find the occupied channel. Thus, $T_p=2K$.

After entering into the piggyback phase, the new user can identify all parameters immediately whenever old user leaves the network. This is because user can leave only when she is master. Else, the new user needs to sense the channel for at least $4K$ time slots. This is because the old user will be silent in the first SB of each MB when she is not a master and MB duration is $2K$ slots. Thus, the duration between consecutive silent SSB is at most $4K-2$ slots. This is due to fact that user transmits in SSB when he is master and he can be master only once in OHS block. Thus new user can sense consecutive transmissions over $2(K-1)$ slots before the old user becomes master and further consecutive transmissions over $2K$ slots when old user\textquoteright s reserved channel is the most preferred channel. Thus, the new user has to wait for one more SB (i.e. SSB for subsequent MB) to identify all block parameters i.e. total $2(K-1)+2K+2=4K$ slots. However, in the worse case, the user may have to wait for $K$ SBs till old user becomes master and hence, she needs to sense the same channel for at most $2K^2$ slots. Then, $T_i=\max(4K,2K^2)=2K^2\quad \forall K>1$.

Next, new user needs to identify the reserved channel before entering into the SMCS phase. As discussed before, the new user sequentially senses the channels and locks on the vacant channel. The worst case corresponds to ($K-1$) active users and master switches to the vacant channel. In this case, the new user needs to sense all channels except reserved channel of a master for at most $2K$ time slots before realizing the master switch and then occupying the master's previous reserved channel in the next time slot. Thus, $T_r=2K+1$. Then,
\begin{equation}
T_s^d = 2K+2K^2+2K+1 = K(2K+4)+1
\end{equation}
This completes the proof. \hfill\IEEEQED

\subsection*{\textbf{Proof of Lemma 3}}
The proof of Lemma 3 is based on Theorem 1. When a new user enters into the network, she has to learn the statistics of ($K-N$) channels which are left unoccupied by existing $N$ users. Thus, using Theorem 1, we have
\[t_m^{nu} \leq \frac{M-1-\sqrt{(M-1)^2- 4M}}{2}, \mbox{  where } M:=\frac{16(K-N)}{\Delta_{\min}^2}.\] 
Thus, for all $t \geq t_m^{nu}$ every switch to one of the ($K-N$) channels results in decrease in the network potential with probability ($1-2t^{-4}$). Thereafter, new user needs only one opportunity to become master and switch to the most preferred channel among ($K-N$) vacant channels. Using the Theorem 1, we have $T_e^d(\delta) = t_m^{nu}+\log \left(\frac{\delta}{1-P_{soc}^{nu}}\right)$ where $P_{soc}^{nu}= 1-2({t_m^{nu}})^{-4}$. \hfill\IEEEQED

\subsection*{\textbf{Proof of Lemma 4}}
When one of the $N$ users leaves the network, each of the remaining users needs at least $N-1$ opportunities to become master and check the feasibility of channel swap or switch to the channel vacated by leaving user. Since the network was in SOC, Theorem 1 guarantees that all users have the sufficient number of samples of each channel and every swap or switch guarantees the decrease of network potential with high probability. Thus, the network will be in SOC again after ($N-1$) OHS blocks, i.e. $2K^2(N-1)$ time slots. Since $N$ is unknown and $N\leq K$, the maximum number of slots required by the network to reach SOC is $T_l^d=2K^2(K-1)$.\hfill\IEEEQED

\vspace{-1cm}
\begin{IEEEbiography}
	[{\includegraphics[width=1in,height=1.25in,clip,keepaspectratio]{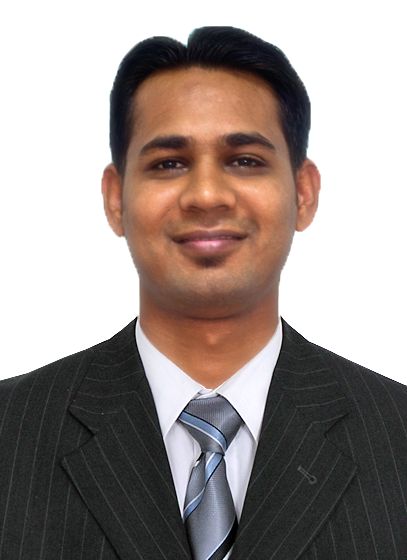}}]{Sumit J. Darak}
	received his Bachelor of Engineering (B.E.) degree in Electronics and Telecommunications Engineering from Pune University, India in 2007, and PhD degree from Nanyang Technological University (NTU), Singapore in 2013.
	He is currently an Assistant Professor at IIIT-Delhi, India. Dr. Sumit has been awarded \textit{DST Inspire Faculty Award} which is a prestigious award for young researchers under 32 years age. He has received \textit{Best Demo Award} at CROWNCOM 2016, \textit{Young Scientist Paper Award} at URSI 2014 and 2017, \textit{Best Student Paper Award} at IEEE DASC 2017 and Second-best poster award in COMSNETs 2019. His current research interests include the reinforcement learning algorithms and reconfigurable architectures for applications such as wireless communications, energy harvesting etc.
	
\end{IEEEbiography}

\begin{IEEEbiography}[{\includegraphics[width=1in,height=2in,clip,keepaspectratio]{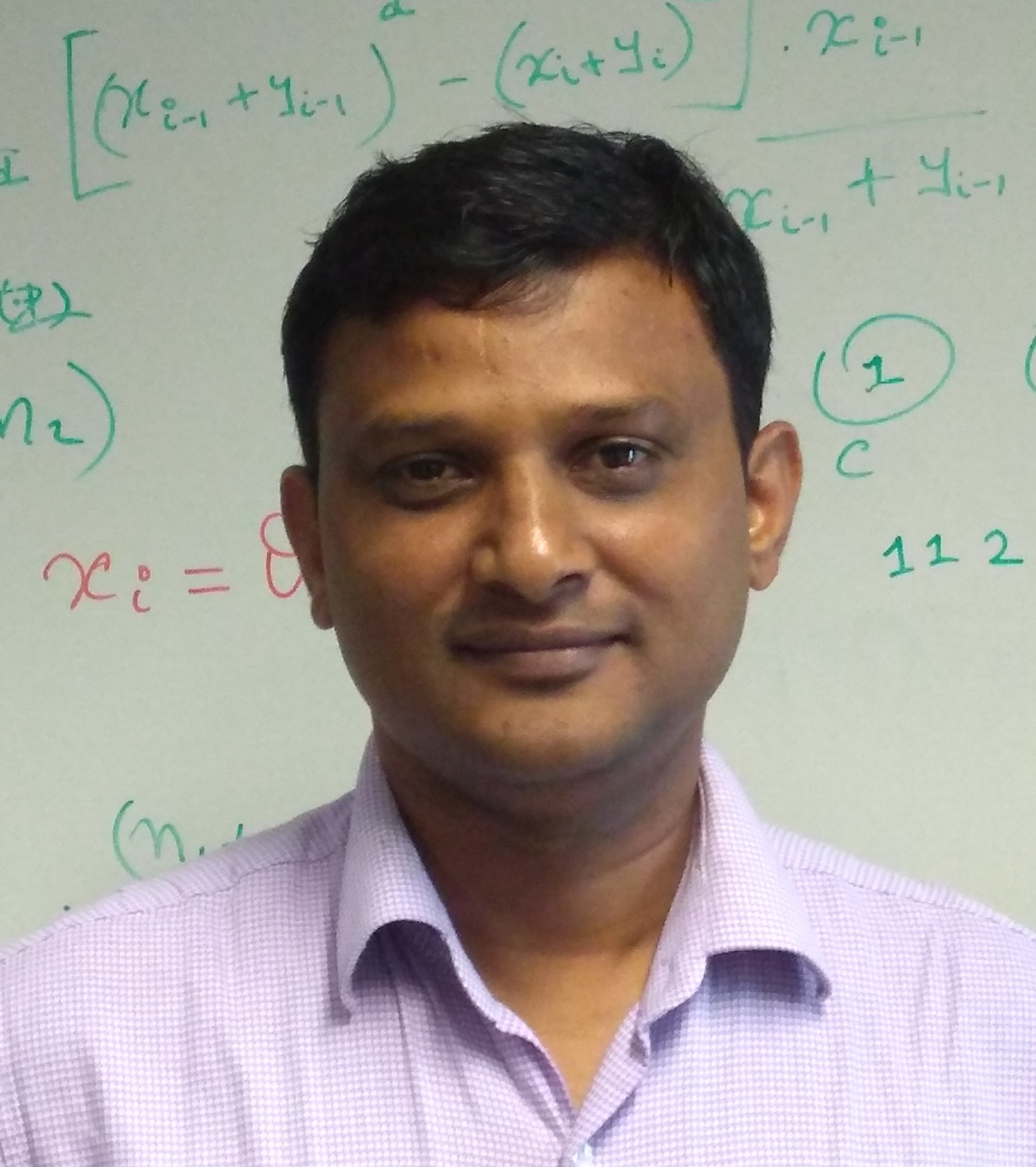}}]{Manjesh K. Hanawal}
	received the M.S. degree in ECE from the Indian Institute of Science, Bangalore, India, in 2009,
	and the Ph.D. degree from INRIA, Sophia Antipolis, France, and the University of Avignon, Avignon, France, in 2013. After two
	years of postdoc at Boston University, he is now an Assistant Professor in Industrial Engineering and Operations Research
	at the Indian Institute of Technology Bombay, Mumbai, India. His research interests include performance evaluation, machine learning
	and network economics. He is a recipient of Inspire Faculty Award from DST and Early Career Research Award from SERB. 
\end{IEEEbiography}

\end{document}